\newif\iflandscape
\newif\ifportrait
\typein[\lorp]%
{Typein "l" (for landscape, twocolumn) or "p" (for portrait, onecolumn)}
\if l\lorp \landscapetrue \else \portraittrue \fi
\newlength{\extralineskip}
\ifportrait
  \documentstyle[12pt]{article}
\fi
\iflandscape
  \documentstyle[twocolumn]{article}
\fi
\makeatletter
\newdimen\normalarrayskip              
\newdimen\minarrayskip                 
\normalarrayskip\baselineskip
\minarrayskip\jot
\newif\ifold             \oldtrue            \def\new{\oldfalse}
\def\arraymode{\ifold\relax\else\displaystyle\fi} 
\def\eqnumphantom{\phantom{(\theequation)}}     
\def\@arrayskip{\ifold\baselineskip\z@\lineskip\z@
     \else
     \baselineskip\minarrayskip\lineskip2\minarrayskip\fi}
\def\@arrayclassz{\ifcase \@lastchclass \@acolampacol \or
\@ampacol \or \or \or \@addamp \or
   \@acolampacol \or \@firstampfalse \@acol \fi
\edef\@preamble{\@preamble
  \ifcase \@chnum
     \hfil$\relax\arraymode\@sharp$\hfil
     \or $\relax\arraymode\@sharp$\hfil
     \or \hfil$\relax\arraymode\@sharp$\fi}}
\def\@array[#1]#2{\setbox\@arstrutbox=\hbox{\vrule
     height\arraystretch \ht\strutbox
     depth\arraystretch \dp\strutbox
     width\z@}\@mkpream{#2}\edef\@preamble{\halign \noexpand\@halignto
\bgroup \tabskip\z@ \@arstrut \@preamble \tabskip\z@ \cr}%
\let\@startpbox\@@startpbox \let\@endpbox\@@endpbox
  \if #1t\vtop \else \if#1b\vbox \else \vcenter \fi\fi
  \bgroup \let\par\relax
  \let\@sharp##\let\protect\relax
  \@arrayskip\@preamble}
\def\eqnarray{\stepcounter{equation}%
              \let\@currentlabel=\theequation
              \global\@eqnswtrue
              \global\@eqcnt\z@
              \tabskip\@centering
              \let\\=\@eqncr
              $$%
 \halign to \displaywidth\bgroup
    \eqnumphantom\@eqnsel\hskip\@centering
    $\displaystyle \tabskip\z@ {##}$%
    &\global\@eqcnt\@ne \hskip 2\arraycolsep
         $\displaystyle\arraymode{##}$\hfil
    &\global\@eqcnt\tw@ \hskip 2\arraycolsep
         $\displaystyle\tabskip\z@{##}$\hfil
         \tabskip\@centering
    &{##}\tabskip\z@\cr}
\makeatother
\def\tr#1{{\rm tr}\kern-3pt\left[#1\right]}

\def\bea{\begin{eqnarray}}
\def\eea{\end{eqnarray}}
\def\bqa{\begin{eqnarray}}
\def\eqa{\end{eqnarray}}

\def\noi{\noindent}
\def\nn{\nonumber}

\def\beq{\begin{equation}}
\def\eeq{\end{equation}}
\def\ba{\beq\new\begin{array}{c}}
\def\ea{\end{array}\eeq}
\def\be{\ba}
\def\ee{\ea}
\def\2{{1\over 2}}
\def\stackreb#1#2{\mathrel{\mathop{#2}\limits_{#1}}}

\def\f{1\over}
\def\Tr{{\rm Tr}}
\def\d{\partial\over\partial}

\renewcommand{\theequation}{\thesection.\arabic{equation}}

\font\teneufm=cmmib10
\font\seveneufm=cmmib7
\font\fiveeufm=cmmib5
\def\bfit#1{{\textfont1=\teneufm\scriptfont1=\seveneufm
\scriptscriptfont1=\fiveeufm
\mathchoice{\hbox{$\displaystyle#1$}}{\hbox{$\textstyle#1$}}
{\hbox{$\scriptstyle#1$}}{\hbox{$\scriptscriptstyle#1$}}}}

\def\balpha{{\bfit\alpha}}
\def\bbeta{{\bfit\beta}}

\def\bmu{{\bfit\mu}}
\def\bxi{{\bfit\xi}}
\def\bphi{{\bfit\phi}}
\def\blambda{{\bfit\lambda}}
\def\brho{{\bfit\rho}}
\def\bj{{\bfit j}}
\def\bfe{{\bfit e}}

\def\bp{{\bfit p}}

\newcommand{\ad}{{\rm ad}}
\newcommand{\Ad}{{\rm Ad}}
\newcommand{\ti}[1]{\tilde{#1}}
\newcommand{\om}{\omega}

\newcommand{\de}{\delta}
\newcommand{\al}{\alpha}

\newcommand{\bet}{\beta}

\newcommand{\La}{\Lambda}
\newcommand{\De}{\Delta}

\newcommand{\vf}{\varphi}
\newcommand{\G}{\Gamma}

\newcommand{\ga}{\gamma}
\newcommand{\li}{\lim_{t\rightarrow \infty}}
\newcommand{\ms}{\mapsto}
\newcommand{\si}{\sigma}
\newcommand{\beql}[1]{\be\label{#1}}
\newcommand{\eq}{\ee}

\newcommand{\p}{\partial}
\newcommand{\di}{{\rm diag}}
\newcommand{\rar}{\rightarrow}

\newcommand{\sm}{\setminus}

\begin{document}
\begin{titlepage}
\setcounter{footnote}0
\begin{center}
\hfill ITEP M4/TH-7/95\\
\hfill FIAN/TD-18/95\\
\hfill hep-th/9601161\\
\vspace{0.3in}
\ifportrait
{\LARGE\bf Liouville Type Models in Group Theory Framework}\\
\vspace{0.15in}
{\Large\bf I. Finite-Dimensional Algebras}
\fi
\iflandscape
{\LARGE\bf Liouville Type Models in Group Theory Framework}\\
\vspace{0.15in}
{\Large\bf I. Finite-Dimensional Algebras}
\\
\fi
\bigskip

\bigskip

{\Large A.Gerasimov\footnote{E-mail address: gerasimov@vxitep.itep.ru
}}$\phantom{hj}^{\dag}$,
{\Large S.Kharchev\footnote{E-mail address:
kharchev@vxitep.itep.ru}}$\phantom{hj}^{\dag}$,
{\Large A. Marshakov\footnote{E-mail address:
mars@lpi.ac.ru}}
$\phantom{hj}^{\ddag,\ \dag}$,\\
{\Large A. Mironov\footnote{E-mail address:
mironov@lpi.ac.ru}}$\phantom{hj}^{\ddag,\ \dag}$,
{\Large A.Morozov\footnote{E-mail address:
morozov@vxitep.itep.ru}}$\phantom{hj}^{\dag}$,
{\Large M.Olshanetsky\footnote{E-mail address:
olshanez@vxdesy.desy.de}}$\phantom{hj}^{\dag}$, \\ \bigskip

\bigskip

\begin{quotation}{

$\phantom{hj}^{\dag}$ --
{\it ITEP, Bol.Cheremushkinskaya, 25, Moscow, 117 259, Russia}\\

$\phantom{hj}^{\ddag}$ --
{\it Theory Department,  P. N. Lebedev Physics
Institute, Leninsky prospect, 53, Moscow,~117924, Russia}}
\end{quotation}
\end{center}
\newpage
\centerline{\bf ABSTRACT}
\begin{quotation}
In the series of papers we represent the ``Whittaker'' wave functional of
$d+1$-dimensional Liouville model as a correlator in
$d+0$-dimensional theory of the sine-Gordon type (for $d=0$ and $1$).
Asypmtotics of this wave function is characterized by the Harish-Chandra
function, which is shown to be a product of simple $\Gamma$-function factors
over all positive roots of the corresponding algebras (finite-dimensional for
$d=0$ and affine for $d=1$). This is in nice correspondence with the recent
results on 2- and 3-point correlators in $1+1$ Liouville model, where emergence
of peculiar double-periodicity is observed.  The Whittaker wave functions of
$d+1$-dimensional non-affine ("conformal") Toda type models are given by simple
averages in the $d+0$ dimensional theories of the affine Toda type.
This phenomenon is in obvious parallel with representation of the free-field
wave functional, which is originally a Gaussian integral over interior of a
$d+1$-dimensional disk with given boundary conditions, as a (non-local)
quadratic integral over the $d$-dimensional boundary itself.  In the present
paper we mostly concentrate on the finite-dimensional case. The
results for finite-dimensional "Iwasawa" Whittaker functions were known, and we
present their survey.  We also construct new "Gauss"  Whittaker functions.
\end{quotation} \end{titlepage}
\clearpage \newpage \tableofcontents \newpage \section{Introduction}
\setcounter{footnote}{0}
\setcounter{equation}{0}
Recent progress in the $2d$ Liouville theory \cite{GerS,DO,ZZ}
encourages one to make an attempt of full value in this long-standing problem
of theoretical physics. It becomes more and more clear that a minor
modification of the free field formalism is needed for the description of the
Liouville correlators -- a statement, long believed in within the framework of
Hamiltonian reduction technique \cite{DS,Kol,Wak,GMMOS}. The crucial point of
this approach is the hidden group structure responsible for dynamical
properties of the system.

Among other, this group structure provides a representation of the wave
function of $d+1$-dimensional {\it integrable} theory in the form of
$d$-dimensional functional integral (time excluded).
Moreover, the asymptotics of the wave function (Harish-Chandra
functions) can be further reduced to exponential $d-1$-dimensional integrals
(products).  These asymptotics are of great importance, since their ratio
determines the $S$-matrix of the theory. Alternatively, $S$-matrix is
proportional to the 2-point function. As for the 3-point function
(form-factor), the quantum mechanics is
certainly not sufficient to fix it, because one needs to introduce additionally
the particle creation vertex. But this is a specifics of the Liouville theory
that the ratio of the two reflected 3-point functions is equal to the 2-point
function \cite{ZZ}.  This property allows one to restore the 3-point amplitude.

Therefore, our main purpose is to construct $d$-dimensional integral
representation for the wave function and calculate its asymptotics.
Technically the problem is a variation of the well-known free-field
representation of Wess-\-Zumino-\-Novikov-\-Witten model and related
coset models, which is in turn related to geometrical quantization
on orbit manifolds (see \cite{GMMOS} for details).

Although the problem of the main interest is the $d+1$-dimensional case,
with $d\ge 1$ ($d=1$ case being described by the affine algebras), we start our
investigation in this paper with the simplest $d=0$ case described by the
finite-dimensional algebras. This case was much studied previously, and the
paper can be partially taken as a survey.  Our other purpose here is to prepare
all the necessary background for the second paper in the series, devoted to the
affine case.

The group
theoretical approach to the Toda theories was considered by many authors
starting
from the first work of O.Bogoyavlensky \cite{Bo}. It was clear from the
very beginning that the open Toda models are related to the simple Lie
algebras, while the periodic Toda models are coming from the affine algebras
(see, for example, review \cite{OP1}). Since an open Toda model
can be derived by the Hamiltonian reduction from the free system on the
cotangent bundle to a simple real Lie group \cite{OP2,Ko2,OP3},
it is natural to suggest that the quantum theory of the model is based on
the irreducible unitary representations of this group. It turned out that
the relevant representations belong to the so-called Whittaker
model  \cite{J,Sch,Ko1,Ha}.
The idea of applying these representations  to the quantization of the open
Toda is due to Kostant (unpublished). It was elaborated in detail later by
Semenov-Tian-Shansky \cite{STSh}. This approach allowed one to give explicit
expressions for the wave functions and to find the $S$-matrix. In the group
theoretical terms the wave functions of the open Toda model are related
in a very simple way to the Whittaker function,
while the $S$ matrix is defined by the Harish-Chandra function. This function
determines the Plancherel measure on the set of
irreducible unitary representations contributing to the Whittaker model.

There are some
interesting results for this class of models, which were obtained beyond
the group-theoretical approach \cite{Gu,Le,Ta,St}, but they dealt with
the two- and three-body problems only.

Of the main interest, however, is the Liouville field theory, which on
the classical level is the reduced system, coming from the cotangent
bundle to the central extended loop group $\widehat{L(SL_2)}$ \cite {AS}.
Therefore, it should be suggested that some analog of the Whittaker
model for $\widehat{L(SL_2)}$ is responsible for the quantization of
the Liouville field theory. More concretely, we present in the second
paper of the series the
group-theoretical derivation of formulas for correlators obtained recently in
\cite{DO,ZZ}. Our approach is very close to the philosophy of "geometrization"
of the scattering in quantum integrable systems advocated by
Freund and Zabrodin \cite{FZ}. They suggested that there should be a
correspondence
between $S$-matrices and Harish-Chandra functions for some groups or
their suitable deformations (quantum, p-adic, elliptic etc.).

As we mentioned above there is another theory --
the quantum periodic Toda lattice --
which is also related to the representations of the affine algebras. From this
point of view, it was considered in the last paper of \cite{GW}. Some
results for the two and three body quantum periodic Toda chains \cite{Gu} have
a similar  interpretation. However, we are not aware of any counterparts of the
Whittaker models for the affine groups.

The Whittaker model was applied to construct automorphic representations
for the groups over the  ring of adels \cite{JL,GKa}.  Later it was
used in \cite{Dr} in the construction of geometric Langlands correspondence
in the case $GL(2)$.

The Whittaker function for the quantum Lorentz group  was defined in \cite{OR}.
It arises there as a wave function in some integrable discretization of the
Liouville quantum mechanics.

Considerable part of the material we consider in detail in this paper, i.e.
related to the finite dimensional case
is well known and can be found in \cite{STSh} and \cite{Ha}.
Our presentation is customized to the extension to the affine case.
We add  also some new results (Gauss-Whittaker function, modification of the
Harish-Chandra function), which will be used in the infinite-dimensional
situation.
Furthermore,  due to a parallel between the quantum open Toda model
and quantum Calogero-Sutherland model, which follows from their group
theoretical origin, we briefly describe  the later model and the corresponding
 spherical model of irreducible representations.
More concretely, the wave functions of the quantum Calogero-Sutherland models
are gauge transform of the zonal spherical functions on a real simple group
$G$.
The later are matrix
elements in the unitary irreducible representations with both the bra and
the ket
vectors being $K$-invariant, where $K$ is a maximal compact subgroup of $G$.
This definition is based on the Cartan decomposition of $G$. The wave functions
for the Toda models
are gauge transform of the Whittaker functions. They are defined as
matrix elements in the same representations with pairing between
$K$-invariant and $N$-covariant states, ($N$ is the positive nilpotent
subgroup)
and are related to the Iwasawa decomposition.
We also consider the pairing between $N$-covariant and $\bar{N}$-covariant
states, ($\bar{N}$ is the negative nilpotent subgroup), which is coming from
the Gauss decomposition. Thereby, we cover all
possible non-trivial reductions to the Cartan subgroups.

The paper organized as follows. First, we discuss the general construction
and point out some important relations with the affine case. Second,
in Part 1 we consider the particular examples of $SL(2)$, $SL(3)$ and
$SL(N)$ in detail. At last, the second Part is devoted to the general
construction.

Throughout the paper we use standard notations and constructions of the
group theory without additional references. One can use, say, \cite{Wa,He}
as standard text-books.

\section{General scheme and comments}
\setcounter{equation}{0}
Let us explain the idea of group theory construction of the wave
functions in the very general terms.

1) Let $g(\xi|T) \in {\cal A}_G(\xi)\otimes {\cal U}_G$ be
the ``universal group element'' of a Lie group $G$ (perhaps, quantum group)
\cite{Mir,MV},
where $\xi$ parametrizes
somehow the group manifold, and $T$ are generators of $G$ in some
(not obligatory irreducible) representation, their only property
being $[T^a, T^b] = f^{abc}T^c$.

2) For every given parametrization $\{\xi\}$, one can introduce two
sets of differential (difference) operators ${\cal D}_{R,L}(\xi)$,
such that
\be
{\cal D}_L^a(\xi) g(\xi|T) = T^a g(\xi|T), \nn \\
{\cal D}_R^a(\xi) g(\xi|T) = g(\xi|T) T^a.
\ee
These operators satisfy the obvious commutation relations:
\be
\left[{\cal D}_L^a, {\cal D}_L^b\right] = -f^{abc}{\cal D}_L^c, \nn \\
\left[{\cal D}_R^a, {\cal D}_R^b\right] = f^{abc}{\cal D}_R^c, \nn \\
\left[{\cal D}_L^a, {\cal D}_R^b\right] = 0.
\label{coreDD}
\ee

3) For given representation ${\cal R}$, scalar product $<\ |\ >$
and two elements $<\psi_L|$ and $|\psi_R>$, one can construct
matrix element
\be
F_{{\cal R}}(\xi | \psi_L,\psi_R) =
<\psi_L | g(\xi|T) |\psi_R>.
\ee
Then, action of any combination of differential operators
${\cal D}_R$ on $F$ inserts the same combination of
generators $T$ to the right of $g(\xi|T)$.
If $|\psi_R>$ happens to be an eigenvector of this combination
of generators, the corresponding $F$ provides a solution to
the differential equation.

4) Of special interest are Casimir operators, since
$|\psi_R>$, which are their eigenvectors can be easily described
as elements of irreducible representations, or linear
combinations of such elements in degenerate cases when different
representations have the same values of Casimir operators.

5) Quadratic Casimir operators, when expressed through
${\cal D}_R$, can serve as Laplace operators, leading to
some important Schr\"odinger equations, of which that for
the Liouville theory is a typical example. In order to obtain the
Liouville model, one should impose additional constraints on the
states $<\psi_L|$ and $|\psi_R>$, what corresponds to the Hamiltonian
 reduction of the free motion on the group
manifold of $G$. Such a reduction is usually associated with some
decomposition of $G$ into a product of subgroups (like Gauss
or Iwasawa decompositions), moreover, different decompositions
can give rise to equivalent reductions. Essentially, reduction
allows one to eliminate the dependence of $F(\xi)$ on some of the coordinates
$\xi$. In the case of finite-dimensional Lie groups, the remaining
coordinates can be made just those on the Cartan torus, while
for affine algebras
it is more interesting to keep the dependence of all the diagonal matrices.

Let us stress that there are, at least, three different ways to obtain the
reduced Hamiltonians. The first one described above is to express
the Casimir operator through $\cal {D}_R$ and impose the reduction
condition after this. However, one can calculate the reduced Hamiltonian
immediately from the matrix element $F_{{\cal R}}(\xi | \psi_L,\psi_R)$
inserting the Casimir operator into $<\psi_L | g(\xi|T) |\psi_R>$. We
illustrate these procedures with concrete examples in Part I. There is also
another way, which we use in Part II, based on the observation that the
second Casimir operator coincides with the Laplace-Beltrami operator, which
can be calculated making use of the Killing metric in the concrete (Iwasawa
or Gauss) coordinates. This way is however inconvenient for constructing
higher Casimir operators.

Now we briefly comment on the general procedure postponing the detailed
discussion till Part II, while illustrative examples are considered in Part I.

First of all, let us remark that the differential operators ${\cal D}_L$
and ${\cal D}_R$ realize respectively the left and the right regular
representations of the algebra of $G$ (in fact, the left one is
anti-representation),
that can be also given by the action on
the space $A_G$ of functions on the group:

\be
\pi_{reg}^L(h)f(g)=f(hg),\ \ \ \pi_{reg}^R(h)f(g)=f(gh),\ \ \ g,h\in G.
\ee
Manifestly these operators can be constructed in the following way. Let us
consider the universal group element $g$ (it is
sufficient to consider $g$ only in the fundamental representation,
since matrix elements in the fundamental representation are generating elements
of the whole algebra $A_G$, and the action of the group can be
extended to the whole algebra $A_G$ making use of comultiplication) and
the (formal) differential operator $\rm d$ acting as the
full derivative on functions of $\xi_i$, i.e. ${\rm d}\equiv \sum_id\xi_i
{\d\xi_i}$. Then, one may calculate, say, $g^{-1}\cdot{\rm d}g$
(Maurer-Cartan form)
and expand it in the generators of the algebra (since
the fundamental representation is sufficient to fix the coefficients
$c_{a,i}$, the calculations are very simple):

\be
g^{-1}\cdot{\rm d}g=\sum_{a,i}c_{a,i}T_ad\xi_i,
\ee
i.e.
\be
{\rm d}g=\sum_{a,i}c_{a,i}\left({\cal D}_R^ag\right)d\xi_i.
\ee
Now one reads off the manifest form of the differential operators
${\cal D}_R$ from this expression. Analogously, one can calculate
${\cal D}_L$.

Our second remark concerns different possible reductions. In fact, there are
three principally different kinds of reductions induced by the Iwasawa and
Gauss
decompositions and by reducing to the radial part of the Cartan decomposition.
The first two reductions lead to the same Liouville wave functions while the
last one -- to the zonal spherical functions. In this paper we are interested
both in the Gauss decomposition and in the Iwasawa one. One of the main points
of our interest are the asymptotics of the Liouville wave function --
Harish-Chandra functions. The number of different asymptotics is equal to the
number of elements of the Weyl group. The ratios of them give the $S$-matrices
and, equivalently, 2-point functions. We demonstrate in this paper that the
Harish-Chandra functions for the finite-dimensional groups are equal to

\be
c(\blambda)=\prod_{\balpha\in\De^+}{\f \Gamma(1+\blambda\cdot\balpha)}
\ee
and all the other are obtained by the action of the Weyl group on $\blambda$.

This form can be easily continued to the affine case.
Indeed, let us consider $\widehat{SL(2)}$ case, for the sake of simplicity.
Then, the system of positive roots is
\be
\balpha_0 + n(\balpha_0+\balpha_1),\
n(\balpha_0+\balpha_1),\
\balpha_1 + n(\balpha_0+\balpha_1),\
n = 0,1,2,\ldots\ .
\ee
Let
\be
\blambda\cdot\balpha_0 = \frac{1}{2} - p + \tau,\ \ \
\blambda\cdot\balpha_1 = -\frac{1}{2} +p.
\ee
Then, one should shift the argument of the $\Gamma$-functions to
1/2 because of the affine situation to obtain
\be\label{HCHA}
c(\blambda) = \prod_{n\geq 0}
\Gamma^{-1}(p + n\tau)
\prod_{n\geq 1}\Gamma^{-1}(n\tau)\Gamma^{-1}(1-p + n\tau).
\ee
This expression still requires a careful regularization,
but all the infinite products cancel from the corresponding reflection
$S$-matrix
(2-point function)\footnote{Throughout the paper we omit from the $S$-matrix
a trivial factor depending on the cosmological constant $\mu$.}
\be
S(p) = \frac{c(-p)}{c(p)} =
\frac{\Gamma(1+p)\Gamma(1+{p\over\tau})}{\Gamma(1-p)\Gamma(1-{p\over\tau})}.
\ee
This expression is to be compared with the formulas for the 2-point
functions obtained in papers \cite{DO,ZZ} in a very different
way\footnote{In notations of \cite{ZZ}, $p=2iP/b$ and $\tau=b^2$.}.
In the second paper of the series we are going to discuss the properties
of the functions of such a type.

Most important, (\ref{HCHA}) exhibits the double-(quasi)periodicity property --
it can be symbolically (modulo requested regularizations) represented as

\be
c(\blambda)\sim\prod_{m,n}(p+m+n\tau).
\ee
From this observation it is just a step to consider the most fundamental object

\be
i(\blambda)=i(p,\tau)\sim\prod_{{\hbox{one}}\atop{\hbox{quadrant}}}(p+m+n\tau).
\ee
Indeed, $i(p)$ is a building block for both the elliptic theta-functions,

\be
\theta(p+\2+{\tau\over 2},\tau)\sim i(p,\tau)i(-p,\tau)i(p,-\tau)i(-p,-\tau)
\ee
and the quantum exponentials,

\be
e_q(e^{2\pi ip})\sim {\f i(p,\tau)i(-p,-\tau)},\ \ \ 1=e^{i\pi\tau}.
\ee

\newpage
\part{Particular groups}
\setcounter{section}{0}
\section{$SL(2)$}
\setcounter{equation}{0}
\subsection{Notations}
The Lie algebra is defined by the relations:
\be
\phantom{dsd}[T_+,T_-]=T_0,\ \ [T_{\pm},T_0]=\mp 2T_{\pm}.
\ee
The fundamental representation:
\be
T_0=\left(\begin{array}{cc}1&0\\0&-1\end{array}\right),\
T_+=\left(\begin{array}{cc}0&1\\0&0\end{array}\right),\
T_-=\left(\begin{array}{cc}0&0\\1&0\end{array}\right).
\ee
The quadratic Casimir operator:
\be
C=(T_-T_++T_+T_-)+\2 T_0^2=2T_-T_++T_0+\2 T_0^2.
\ee
Different parametrizations of the group element:
\be
g_I(\theta,\phi,\chi|T) = e^{\theta T_2}e^{\phi_I T_0}e^{\chi_I T_+},\\
g_G(\psi,\phi,\chi|T) = e^{\psi T_-}e^{\phi T_0}e^{\chi T_+}.
\ee
Group element in the fundamental representation:
\be\label{groupelI}
g_I = \left( \begin{array}{cc} \cos\theta & \sin\theta \\
                  -\sin\theta & \cos\theta \end{array}\right)
 \left( \begin{array}{cc} e^{\phi_I} & 0 \\ 0 & e^{-\phi_I} \end{array}\right)
   \left( \begin{array}{cc} 1 & \chi_I \\ 0 & 1 \end{array}\right) = \nn\\
  = \left( \begin{array}{cc} e^{\phi_I}\cos\theta &
                   \chi_I e^{\phi_I}\cos\theta + e^{-\phi_I}\sin\theta \\
 -e^{\phi_I}\sin\theta & -\chi_I e^{\phi_I}\sin\theta + e^{-\phi_I}\cos\theta
             \end{array}\right),
\ee
\be\label{groupelG}
g_G = \left( \begin{array}{cc} 1 & 0 \\ \psi & 1 \end{array}\right)
 \left( \begin{array}{cc} e^{\phi} & 0 \\ 0 & e^{-\phi} \end{array}\right)
   \left( \begin{array}{cc} 1 & \chi \\ 0 & 1 \end{array}\right) =
  \left( \begin{array}{cc} e^{\phi} & \chi e^{\phi} \\
   \psi e^{\phi} & \psi\chi e^{\phi} + e^{-\phi} \end{array}\right).
\ee
Connection between different parametrizations:
\be
\psi = -\tan\theta, \nn\\
e^\phi = e^{\phi_I}\cos\theta, \nn\\
\chi = \chi_I + e^{-2\phi_I}\tan\theta.
\ee

\subsection{Representations}
\subsubsection{Regular representations}
Right and left regular representations can be read off using the
manifest expressions for the "currents" $g^{-1}\cdot {\rm d}g$ and
${\rm d}g\cdot g^{-1}$. In the Iwasawa case, these are
\be\label{gdgI}
g^{-1}_Idg_I = \left(\begin{array}{cc}
       d\phi_I + e^{2\phi_I}\chi_Id\theta &
    d\chi_I + 2\chi_Id\phi_I + (\chi_I^2e^{2\phi_I}+e^{-2\phi_I})d\theta \\
      -e^{2\phi_I}d\theta & - d\phi_I - e^{2\phi_I}\chi_Id\theta
                   \end{array}\right), \nn \\
dg_Ig^{-1}_I = \left(\begin{array}{cc}
      \cos 2\theta d\phi_I + \2 e^{2\phi_I}\sin 2\theta d\chi_I &
 e^{2\phi_I}\cos^2\theta d\chi_I - \sin 2\theta d\phi_I + d\theta \\
 -e^{2\phi_I}\sin^2\theta d\chi_I - \sin 2\theta d\phi_I - d\theta
      & -\cos 2\theta d\phi_I - \2 e^{2\phi_I}
\sin 2\theta d\chi_I \end{array}\right),
\ee
while, for the Gauss case, one obtains
\be\label{gdgG}
g^{-1}dg = \left(\begin{array}{cc}
      - e^{2\phi}\chi d\psi + d\phi &
              - e^{2\phi}\chi^2 d\psi + 2\chi d\phi + d\chi \\
      e^{2\phi}d\psi & e^{2\phi}\chi d\psi - d\phi \end{array}\right), \\
dg\cdot g^{-1} =  \left(\begin{array}{cc}
      - e^{2\phi}\psi d\chi + d\phi &  e^{2\phi}d\chi \\
 -e^{2\phi}\psi^2 d\chi + 2\psi d\phi + d\psi & e^{2\phi}\psi d\chi - d\phi
         \end{array} \right).
\ee
Using these formulas, one can easily read off the following expressions
\be\label{rrI}
{\partial g_I\over \partial\phi_I}=g(T_0+2\chi_IT_+)=
(\cos 2\theta T_0-\sin 2\theta T_+-\sin 2\theta T_-)g,\\
{\partial g_I\over\partial\theta} = g\left(e^{2\phi_I}\chi_IT_0+
(\chi^2_Ie^{2\phi_I}+e^{-2\phi_I})T_+-e^{2\phi_I}T_-\right)=
\left(T_+-T_-\right)g,\\
{\partial g_I\over\partial\chi_I}=gT_+=\left(\2 e^{2\phi_I}\sin 2\theta
T_0+e^{2\phi_I}\cos^2\theta T_+-e^{2\phi_I}\sin^2\theta T_-\right)g
\ee
and
\be\label{rrG}
\frac{\partial g}{\partial\phi} = g(T^0 + 2\chi T^+) =
                (T^0 + 2\psi T^-)g, \nn \\
\frac{\partial g}{\partial\chi} = gT^+ =
   e^{2\phi}(T^+ - \psi T^0 - \psi^2 T^-)g, \nn \\
\frac{\partial g}{\partial\psi} =
   g(-\chi^2 T^+ - \chi T^0 + T^-)e^{2\phi} = T^-g.
\ee

\noindent
\underline{{\bf Right regular representation}}

\noi
In the Iwasawa case, one gets using (\ref{rrI})
\be\label{rrrI}
{\cal D}_R^+ = \frac{\partial}{\partial\chi_I},\
{\cal D}_R^0 = \frac{\partial}{\partial\phi_I} -
            2\chi_I\frac{\partial}{\partial\chi_I},\\
{\cal D}_R^- = \left(e^{-4\phi_I}-
       \chi_I^2\right)\frac{\partial}{\partial\chi_I} + \chi_I
       \frac{\partial}{\partial\phi_I} -
       e^{-2\phi_I}\frac{\partial}{\partial\theta}.
\ee
In the Gauss case, one should use (\ref{rrG})
\be\label{rrrG}
{\cal D}_R^+ = \frac{\partial}{\partial\chi},\ {\cal D}_R^0=-2\chi{\d\chi}+
{\d\phi},\ {\cal D}_R^-=e^{-2\phi}{\d\psi}+\chi{\d\phi}-\chi^2{\d\chi}.
\ee

\noi
\underline{{\bf Left regular representation}}

\noi
Analogously to the previous paragraph, one gets in the Iwasawa case
\be\label{lrrI}
{\cal D}_L^-=\cos 2\theta e^{-2\phi_I}{\d \chi}-\2\sin 2\theta{\d \phi_I}-
\cos^2\theta{\d \theta},\\
{\cal D}_L^0=2\sin 2\theta e^{-2\phi_I}{\d\chi}+\cos 2\theta {\d \phi_I}-
\sin 2\theta {\d \theta},\\
{\cal D}_L^+=\cos 2\theta e^{-2\phi_I}{\d\chi}-\2\sin 2\theta {\d\phi_I}
+\sin^2\theta {\d\theta}.
\ee
In the Gauss case
\be\label{lrrG}
{\cal D}_L^- = \frac{\partial}{\partial\psi},\ {\cal D}_L^0=-2\psi{\d\psi}
+{\d\phi},\ {\cal D}_L^+=-\psi^2{\d\psi}+\psi{\d\phi}+e^{-2\phi}{\d\chi}.
\ee

\subsubsection{Highest weight representation}
We consider the principal (spherical) series of
representations, induced by the
one-dimensional representations of the Borel subgroup.
The space of representation is functions of one real variable $x$
and matrix elements are defined by integrals with the flat measure.
The action of the group is given by differential operators:
\be\label{j-rep}
T_+ = \frac{\partial}{\partial x},\ T_0=-2x{\d x}
+2j,\ T_-=-x^2{\d x}+2jx.
\ee
In this paper we consider only unitary
highest weight irreducible representations.
The requirement of unitarity
is crucial for our needs since it makes all
integrals convergent. For the series of the representations that we consider
in this paper,
unitarity implies that $j+\2$ is pure imaginary.
Analogously, in the case of general $SL(N)$,
only pure imaginary values of $j_i+\2$ (i.e. $\bj+\brho$) are admissible.

\subsection{Hamiltonian}
\subsubsection{Reducing from the regular representation}
Calculating either in the left regular representation, or in the right one,
we obtain for the quadratic Casimir operator
in the Iwasawa case
\be
C=\2 {\partial^2\over\partial\phi_I^2}+{\d\phi_I}+2e^{-4\phi_I}
{\partial^2\over\partial \chi_I^2}-2e^{-2\phi_I}{\partial^2\over
\partial\chi_I\partial\theta}=\\=e^{-\phi}\left(
\2 {\partial^2\over\partial\phi_I^2}+2e^{-4\phi_I}
{\partial^2\over\partial \chi_I^2}-2e^{-2\phi_I}{\partial^2\over
\partial\chi_I\partial\theta} - \2 \right)e^{\phi}.
\ee
Its eigenvalue in the spin-$j$ representation is $2(j^2+j)$. Therefore, the
matrix element $F_I=<\psi_L|g(\theta,\phi_I,\chi_I)|\psi_R>$ with $<\psi_L|$
and $\psi_R|$ belonging to the spin-$j$ representation satisfies the equation
\be
CF_I=2(j^2+j)F_I.
\ee
We choose matrix element $F_I$ to lie in the (highest weight)
spin-$j$ (irreducible)
representation of the principal (spherical) series of $SL(2,{\bf R})$.
The Liouville Hamiltonian is obtained from this Casimir operator
after the reduction is imposed
\be
{\d\chi_I}F_I=i\mu F_I,\ {\d\theta}F_I=0,
\ee
which implies (see (\ref{rrrI}) and (\ref{lrrI}))
\be\label{statecondI}
T_+|\psi_R>=i\mu|\psi_R>,\ \ <\psi_L|T_2=0
\ \ (\hbox{i.e.  }<\psi_L|T_-=<\psi_L|T_+).
\ee
Then, the Hamiltonian is equal to
\be\label{HI2}
H=\2 {\partial^2\over\partial\phi_I^2}+{\d\phi_I}-2\mu^2e^{-4\phi_I}
\ee
and the function $\Psi_I(\phi_I)=e^{\phi_I}F_I$ satisfies the following
Schr\"odinger equation
\be\label{SchI}
\left[\2 {\partial^2\over\partial\phi_I^2}-2\mu^2e^{-4\phi_I}\right]
\Psi_I(\phi_I)=2\left(j+{\f 2}\right)^2\Psi_I(\phi_I).
\ee

\bigskip

In the Gauss case the quadratic Casimir operator is
\be
C=\2 {\partial^2\over\partial\phi^2}+{\d\phi}+2e^{-2\phi}
{\partial^2\over\partial \psi\partial\chi}.
\ee
Then, the reduction conditions are
\be {\d\chi}F_G=i\mu_RF_G,\ {\d\psi}F_G=i\mu_LF_G,
\ee
i.e. (see (\ref{rrrG}) and (\ref{lrrG}))
\be\label{statecondG}
T_+|\psi_R>=i\mu_R|\psi_R>,\ \ <\psi_L|T_-=i\mu_L<\psi_L|.
\ee
The Hamiltonian is equal to
\be\label{HG2}
H=\2 {\partial^2\over\partial\phi^2}+{\d\phi}-2\mu_R\mu_Le^{-2\phi}
\ee
and the function $\Psi_G(\phi)=e^{\phi}F_G$
satisfies the following Schr\"odinger equation
\be\label{SchG}
\left[\2
{\partial^2\over\partial\phi^2}-2\mu_R\mu_Le^{-2\phi}\right]
\Psi_G(\phi)=2\left(j+{\f 2}\right)^2\Psi_G(\phi),
\ee

\underline{Comment.} Two Schr\"odinger equations (\ref{SchI}) and
       (\ref{SchG}) are related by the replace $j+{\f 2}\to 2j+1$ and the
appropriate rescalings of $\mu$'s and $\phi$'s.

\subsubsection{Immediate matrix element reduction}
Using the representation (\ref{groupelI}) and conditions (\ref{statecondI}),
we can immediately
obtain Liouville equation with Hamiltonian (\ref{HI2}) in the Iwasawa
case\footnote{From
now on, we consider here the matrix element of $e^{\phi_I T_0}$,
instead of $g$, in order to exclude
the trivial $\chi_I$-dependence of $F_I$, and, analogously,
the $\psi$- and $\chi$-dependencies of $F_G$ below.}
\be
2j(j+1)F_I^{(j)}\equiv 2j(j+1)<\psi_L|e^{\phi_I T_0}|\psi_R>_j=
<\psi_L|e^{\theta T_-}e^{\phi_IT_0}e^{\chi_IT_+}\hat C|\psi_R>_j=\\=
<\psi_L|e^{\phi_IT_0}(2T_-T_++T_0+\2 T_0^2)|\psi_R>_j=\\=
2i\mu e^{-2\phi_I}<\psi_L|T_-e^{\phi_IT_0}T_+|\psi_R>_j+
\left({\d\phi_I}+\2 {\partial^2\over\partial
\phi^2_I}\right)<\psi_L|e^{\phi_I T_0}|\psi_R>_j=\\=
\left(-2\mu^2e^{-4\phi_I}+{\d\phi_I}+\2 {\partial^2\over\partial
\phi^2_I}\right)<\psi_L|e^{\phi_I T_0}|\psi_R>_j.
\ee
Analogously, in the Gauss case, we get from
(\ref{groupelG}) and (\ref{statecondG}) the equation with
Hamiltonian (\ref{HG2})
\be
2j(j+1)F_G^{(j)}\equiv 2j(j+1)<\psi_L|e^{\phi T_0}|\psi_R>_j=
<\psi_L|e^{\psi T_-}e^{\phi T_0}e^{\chi T_+}\hat C|\psi_R>_j=\\=
<\psi_L|e^{\phi T_0}(2T_-T_++T_0+\2 T_0^2)|\psi_R>_j=\\=
\left(-2\mu_R\mu_Le^{-2\phi}+{\d\phi}+\2 {\partial^2\over\partial
\phi^2}\right)<\psi_L|e^{\phi T_0}|\psi_R>_j.
\ee

\subsection{Liouville wave function (LWF) and its asymptotics}
\subsubsection{Solving Liouville equation}
\underline{{\bf Solution}}

\noi
We are looking for the handbook solution to the Liouville equation
\be\label{Le}
{\partial^2 f\over \partial \varphi^2} - \mu^2 e^{2\varphi}f=\lambda^2f.
\ee
The solution is a cylindric function \cite{GR}, which we choose to be
the Macdonald function because of quantum mechanical boundary conditions
($f$ is to be restricted function at pure imaginary $\lambda$)
\be\label{solution}
f=K_\lambda(\mu e^{\varphi}).
\ee

\noi
\underline{{\bf Asymptotics}}

\noi
Now let us investigate the asymptotics
of this solution. Function (\ref{solution}) can be represented in the form
which allows one to distinguish easily two different asymptotical exponentials
at large negative values of
$\varphi$:\footnote{Let us emphasize that, in the Liouville
theory, the oscillating wave exists only at one infinity -- due to the
infinitely
increasing potential at the other one.
Therefore, we should extract two exponentials $e^{\pm \lambda
\varphi}$ at the same
asymptotical region. Technically, this is done as follows:
one puts $\lambda$ real
and takes it positive and negative respectively to extract
appropriate exponentials.
Let us remind that the true value of $\lambda$ is pure imaginary.}

\be\label{as}
K_\lambda(\mu e^{\varphi})
=-{\pi\over 2}{I_\lambda(\mu e^{\varphi})-I_{-\lambda}(\mu e^{\varphi})
\over\sin\pi\lambda}\stackreb{\varphi\to -\infty}{\sim}\\
\stackreb{\varphi\to -\infty}{\sim}
-{\pi\over\sin\pi\lambda}\left\{{\f \Gamma(1+\lambda)}\left({\mu
e^\varphi\over 2}
\right)^\lambda - {\f \Gamma(1- \lambda)}\left({\mu e^\varphi\over 2}
\right)^{-\lambda}\right\},
\ee
where $I_\lambda(z)$ is the Infeld function (the Bessel function of
pure imaginary argument).

Let us derive this asymptotical behaviour using integral representation for the
Macdonald function:
\be\label{as1}
K_\lambda(\mu e^\varphi)={\Gamma(\lambda+{\f 2})\over\Gamma({\f 2})}
\left(\mu e^\varphi\right)^\lambda\int_0^\infty {\cos t\ dt\over
\left(t^2+\mu^2 e^{2\varphi}\right)^{\lambda+{\f 2}}}\longrightarrow\\
\longrightarrow {\Gamma(\lambda+{\f 2})\over\Gamma({\f 2})}
\left(\mu e^\varphi\right)^\lambda\int_0^\infty t^{-2\lambda-1}\cos t\ dt=
-{\pi \over 2\Gamma(\lambda+1)\sin\pi\lambda}\left({\mu\over 4}\right)^\lambda
e^{\lambda \varphi}.
\ee
Here the $\Gamma$-function integral is defined whenever $\lambda\le 0$.
If $\lambda\ge 0$, one can redefine variable $t\to\mu e^\varphi t$ to get
instead of the second line in (\ref{as1}) the second exponential in
(\ref{as})
\be\label{as2}
{\Gamma(\lambda+{\f 2})\over\Gamma({\f 2})}
\left({2\over\mu e^\varphi}\right)^\lambda\int_0^\infty (t^2+1)^{-\lambda-{\f
2}}\ dt
={\Gamma (\lambda)\over 2}\left({2\over\mu}\right)^\lambda e^{-\lambda
\varphi}.
\ee

\subsubsection{Iwasawa Whittaker LWF}
In what follows we reproduce solution (\ref{solution}) to the Liouville
equation following the group theory line of ss.1.2-1.3 (and also using some
direct methods), which is important beyond $SL(2)$ and especially for the
affine algebras.

To calculate matrix element which solves Liouville equation (\ref{SchI}),
we need the manifest solution to conditions (\ref{statecondI}) in the space
of spin-$j$ representation (see sect.1.2.2). We use formulas (\ref{j-rep}).
The first condition in (\ref{statecondI})
\be
T_+ \psi_R={\d x}\psi_R(x)=i\mu\psi_R(x)
\ee
has the evident solution
\be
\psi_R(x)=e^{i\mu x}.
\ee
Writing down the second condition, one needs to take into account that the
differential operator from (\ref{j-rep}) acts to the left state. Then, this
condition is
\be\label{l}
(T_--T_+)\psi_L=(2x+(1+x^2){\d x}+2jx)\psi_L(x)=0
\ee
and the solution to this equation is
\be
\psi_L(x)=(1+x^2)^{-j-1}.
\ee
Therefore, we get finally
\be\label{LWFI}
F^{(j)}_I(\phi_I)=<\psi_L|e^{\phi_IT_0}|\psi_R>_j=
\int_{-\infty}^\infty (1+x^2)^{-j-1} e^{\phi_I(2j-2x{\d x})}e^{i\mu x}\ dx=\\
=e^{2j\phi_I}\int_{-\infty}^\infty (1+x^2)^{-j-1} e^{i\mu xe^{-2\phi_I}}
\ dx\sim
e^{-\phi_I}K_{j+{\f 2}}(\mu e^{-2\phi_I}).
\ee
Since the solution to equation (\ref{SchI}) is $e^{\phi_I}F^{(j)}_I(\phi_I)
=K_{j+{\f 2}}(\mu e^{-2\phi_I})$, it
indeed coincides with (\ref{solution}).

In the course of the calculation, we used in (\ref{l}) the action of
the generators $T_+$ and $T_-$ to the left state, i.e. the corresponding
$x$-derivatives acted to the left. The other way of doing is to act by the same
generators to the right state. It gives the equation

\be
(T_+-T_-)\psi_L=((1+x^2){\d x}-2jx)\psi_L'(x)=0,
\ee
the solution being
\be
\psi_L'(x)=(1+x^2)^{j}.
\ee
Now, since $\lambda=j+\2$ is pure imaginary, one can write

\be
\psi_L(x)=\overline{\psi_L'(x)},\ \ \ \hbox{i.e.}\ \ \
\psi_{L,\lambda}(x)=\psi_{L,\bar\lambda}'(x)=\psi_{L,-\lambda}'(x).
\ee
Such a way of doing is correct in general, since the corresponding
integral (scalar product) exists because of unitarity of the representation
(i.e. pure imaginary $\lambda$ -- see sect.3.2 and Part II).

\subsubsection{Gauss Whittaker LWF}
\underline{{\bf Solution}}

\noi
Similarly to the previous subsection, conditions (\ref{statecondG}) have
the form in the Gauss case:
\be
T_+|\psi_R>={\d x}\psi_R(x)=i\mu_R\psi_R(x),\\
<\psi_L|T_-=(2x+x^2{\d x}+2jx)\psi_L(x)=i\mu_L\psi_L(x).
\ee
The solutions to these conditions are
\be
\psi_R(x)=e^{i\mu_Rx},\ \ \psi_L(x)=x^{-2(j+1)}e^{-{i\mu_L\over x}}.
\ee
Therefore, we get the solution to equation (\ref{SchG})
\be\label{LWFG}
e^{\phi}F_G^{(j)}(\phi)=e^{\phi}<\psi_L|e^{\phi T_0}|\psi_R>_j=\\
=e^{\phi}\int x^{-2(j+1)}e^{-{i\mu_L\over x}}e^{\phi(2j-2x{\d x})}
e^{i\mu_Rx}\ dx=\\= \left({i\over \mu_R}\right)^{-(2j+1)}e^{-(2j+1)\phi}
\int_0^{\infty} x^{-2(j+1)}e^{-{\mu_L\mu_Re^{-2\phi}
\over x}- x}dx=\\=2\left(i\sqrt{{\mu_L\over\mu_R}}\right)^{-(2j+1)}
K_{2j+1}(2\sqrt{\mu_L\mu_R}e^{-\phi})\ dx.
\ee
At the last stage, we have used some different integral representation
(\ref{AMac})
for the Macdonald function. Let us stress that
this expression is a solution to Liouville Schr\"odinger equation (\ref{SchG})
for any choices of integration domain, say, for
$\int_{-\infty}^{+\infty}dx$ or $\int_0^\infty dx$.
The integral converges for the second choice. Then, the change of
variable, $x = e^\xi$, is admissible, which brings it to the form of
the $0+0$-dimensional sine-Gordon model:
\be
\int_{-\infty}^{+\infty}
d\xi \exp\left(-(2j+1)\xi - \mu_Le^{-\xi} - \mu_Re^{\xi -2 \phi}\right).
\ee

\noi
\underline{{\bf Asymptotics}}

\noi
For future use, we derive here the asymptotics of the Macdonald function
starting with this different integral representation (\ref{AMac}):
\be\label{Mac2}
K_\nu(z)={\f 2}\left({z\over 2}\right)^\nu\int_{0}^\infty
e^{-t-{z^2\over 4t}}t^{-\nu-1}dt.
\ee
This integral can be dealt with in
complete analogy with (\ref{as1})-(\ref{as2}).
If $\nu\le 0$, the term ${z^2\over 4t}$ can be thrown away in asymptotics
(i.e. at small z) and the integral is equal to
\be
K_\nu(z)\stackreb{z\to 0}{\sim}{\f 2}\left({z\over
2}\right)^\nu\int_0^\infty e^{-t}t^{-\nu-1}dt=
-{\pi\over 2}{1\over \sin\pi\nu \Gamma(1+\nu)}\left({z\over 2}\right)^\nu.
\ee
On the other hand, if
$\nu\ge 0$, one should do the replace $t\to z^2t$, and the result is
\be
K_\nu(z)\stackreb{z\to 0}{\sim}{\f 2}(2z)^{-\nu}\int_0^\infty
e^{-{1\over 4t}}t^{-\nu-1}dt=
{\pi\over 2}{1\over \sin\pi\nu \Gamma(1-\nu)}\left({z\over 2}\right)^{-\nu}.
\ee
Both these results coincide with formula (\ref{as}).

\subsubsection{Liouville solution by the Fourier transform}
Here we show how the Liouville equation can be solved immediately using
the Fourier transform (see also \cite{Le}).
This method looks the most direct and transparent in all
applications, although we have not worked out it yet in full detail for rank
$>2$ group.

To begin with, we propose the Fourier transform (indeed, with our definition in
this subsection, this is the inverse Fourier transform, which makes
intermediate
formulas more observable) of Liouville equation (\ref{Le}).  Then, the
equation takes the form

\be\label{Lep}
\left(-{\f 4}p^2-e^{2x}\right)\bar f(p)=\lambda^2 \bar f(p),\ \ f(x)=
\int_{-\infty}^{\infty} dp \exp\{ipx\}\bar f(p)
\ee
where, for the sake of simplicity, we put $\mu=1$ and rescale
the coefficient of the first term to be ${\f 4}$.
Now there are two ways to solve this equation. The most immediate one is to
rewrite the operator $e^{2x}$ as a difference operator and to solve the
difference equation obtained. Indeed, this operator acts on functions of $p$ as
shift operator $e^{2x}\bar f(p)=\bar f(p+2i)$. Therefore, equation (\ref{Lep})
can be rewritten in the difference form
\be\label{deq}
\bar
f(p+2i)=-(\lambda^2+{p^2\over 4})\bar f(p).
\ee
Its obvious solution is
\be\label{sl2F}
\bar
f(p)=\Gamma ({p\over 2i}+\lambda)\Gamma ({p\over 2i}-\lambda).
\ee
This
solution can be multiplied by a periodic function still satisfying equation
(\ref{deq})\footnote{This is
amazing that, after the inverse Fourier transform, the arbitrary periodic
function results into only two linearly independent solutions, i.e.
the transformation has a huge kernel.}.
However, the choice of Macdonald solution to equation (\ref{Le})
above is consistent with the purely $\Gamma$-function solution.
Making the inverse Fourier transform, one obtains
\be\label{2G}
f(\varphi)=\int dp e^{ip\varphi} \Gamma
({p\over 2i}+\lambda)\Gamma ({p\over 2i}-\lambda)=\\=
\int_0^{\infty}dx\int_0^{\infty}d\tilde x\int_{-\infty}^{\infty}dp
e^{-x-\tilde x}
x^{{p\over 2i}-\lambda-1}\tilde x^{{p\over 2i}+\lambda-1}e^{ip\varphi}=\\
=\int_0^{\infty}{dx\over x}\int_0^{\infty}
{d\tilde x\over \tilde x}\left\{\int_{-\infty}^{\infty}dp
\left(x\tilde xe^{-2\varphi}\right)^{{p\over 2i}}\right\}e^{-x-\tilde x}
\left({\tilde x\over x}\right)^\lambda
\sim\\\sim
e^{2\lambda
\varphi}\int_0^{\infty}dxx^{-2\lambda-1}e^{-x-{e^{2\varphi}\over x}}
\sim K_{2\lambda}(2e^\varphi).
\ee
Integrations are performed here in the
indicated order so that the first integration over $p$ leads to the
$\delta$-function $\delta(\log (x\tilde x) - 2\varphi)$.
Let us note that this form of the solution is very
convenient to get its asymptotics. Indeed, since the singularities of integral
(\ref{2G}) lie on the imaginary axis, one can close the integration contour
at infinity and calculate the integral using Cauchy theorem. Then, the sum of
the pole contribution gives the expansion at small $z=2e^x$, the leading
asymptotics is given by the nearest singularity of the integral. Depending on
the sign of $\lambda$, these are $p_s=\pm 2i\lambda$. Then, integral (\ref{2G})
is nothing but the value in this pole:  \be f(x)=\int dp e^{ipx} \Gamma
({p\over 2i}+\lambda)\Gamma ({p\over 2i}-\lambda)\stackreb{z\to 0}{\sim}
\Gamma(\pm 2\lambda)z^{\mp\lambda}.  \ee

\subsubsection{Liouville solution by canonical transformation}
Close to the considered in the previous subsection is the following way
of solving the Liouville equation (see also \cite{St}).
Liouville equation (\ref{Lep})
can be considered as the eigenvalue problem of
the Hamiltonian
\be
\hat H = -{\f 4}\hat p^2 - e^{2\varphi}, \ \ \ [\hat p,\hat \varphi] = i
\ee
so that the wave function $f_{\lambda}(\varphi)$ satisfies the equation

\be
(\hat H - \lambda^2)f_\lambda(\varphi) = 0.
\ee
One can make the change of canonical variables $\hat\varphi\to \hat
Q=e^{\hat\varphi}$, $\hat p\to \hat P=e^{-\hat \varphi}(\hat p+2i\lambda)$,
$[\hat P,\hat Q]=i$.  Then, in new variables, the eigenvalue problem simplifies

\be
\hat H - \lambda^2  = -{\f 4}e^{\hat\varphi}\left(
e^{-\hat\varphi}(\hat p+2i\lambda)^2 -
4i\lambda e^{-\hat\varphi}(\hat p+2i\lambda) +4 e^{\hat\varphi}\right) =\\=
-{\f 4}\hat Q\left((\hat P^2+4)\hat Q - i(4\lambda + 1)\hat P\right).
\label{HcanQM}
\ee
Now, choosing $P$-representation, we obtain the non-trivial
zero mode of operator (\ref{HcanQM})

\be
\bar f_{\lambda}(P) =
(P^2 + 4)^{-2\lambda - \frac{1}{2}}.
\ee
Coming back to functions on the original coordinate space (i.e. performing the
inverse Fourier transform), we get finally

\be
f_{\lambda}(\varphi) =e^{-2\lambda \varphi} \int_{-\infty}^{\infty}dP
(P^2 + 4)^{-2\lambda - \frac{1}{2}}e^{-iPe^\varphi} \sim K_{2\lambda}
(2e^\varphi).
\ee
This integral representation coincides with that obtained in the Iwasawa case.
Quite analogically, one can use a different canonical transformation to get the
Gauss integral representation.  Indeed, introducing $\hat Q=e^{2\hat\varphi}$,
$\hat P=\2 e^{-2\hat\varphi}(\hat p+2i\lambda)$ ($[\hat P,\hat Q]=i$), one gets

\be
H - \lambda^2  = -{\f 4}e^{2\varphi}\left(e^{-2\varphi}(p+2i\lambda)^2 -
4i\lambda e^{-2\varphi}(p+2i\lambda) + 4\right) =\\=
-Q\left(P^2Q - i(2\lambda + 1)P+1\right).
\label{HcanQM2}
\ee
Now in $P$-representation the non-trivial
zero mode of operator (\ref{HcanQM2}) is

\be
\bar f_{\lambda}(P) =
P^{-2\lambda - 1}e^{{i\over P}},
\ee
and after the inverse Fourier transform, we get finally

\be
f_{\lambda}(\varphi) =
e^{2\lambda \varphi} \int dP P^{-2\lambda - 1}e^{-{1\over P}-Pe^{2\varphi}}\sim
K_{2\lambda}(2e^\varphi).
\ee

\subsubsection{Harish-Chandra functions}
Now let us say some words about the above-obtained asymptotics of the LWF.
Each separate asymptotics is defined ambiguously, up to
overall normalization factor. This means that only the ratio of the
coefficients in front of oscillating waves makes an invariant sense. Indeed,
this ratio is nothing but reflection $S$-matrix. However, one can impose some
invariant conditions to fix the normalization. Let us require that
the asymptotics have no poles at finite momenta (this requirement still
fix the normalization
ambiguously\footnote{One way to fix this normalization completely is to require
that the leading
large-$p$ asymptotics of the Fourier transformed WF does not depend on the
energy
$\lambda$. Say, in the $SL(2)$ case, formula (\ref{sl2F}) implies that this
asymptotics is $log\bar f(p)
\stackreb{p\to\infty}{\sim}\left({p\over 2i}-1\right)\log
{p\over 2i}-{p\over i}+o(1)$.}).
For the $SL(2)$ Liouville solutions this means that the
asymptotics $\Gamma (\pm \lambda)$ (see (\ref{LWFI}) and (\ref{LWFG}) --
$\lambda$ is pure imaginary and is equal to $j+\2$ for the Iwasawa case and to
$2j+1$ for the Gauss one) do not correspond to the properly normalized LWF but
should be additionally multiplied by the function $\sin\pi\lambda$ which
eliminates all the poles while bringing additional zeros. The two asymptotics
of the so normalized LWF are

\be\label{hchsl2}
c_{\pm}={\f \Gamma (1\pm \lambda)}.
\ee
These functions are
called Harish-Chandra functions \cite{HC} will be discussed for other groups
below.

\section{$SL(3)$}
\setcounter{equation}{0}
\subsection{Notations}
Algebra is completely described by the list of non-vanishing simple root
commutation relations
\be
\phantom{dsd}[T_{+i},T_{-i}]=T_{0,i},\ \
[T_{\pm,i},T_{0,i}]=\mp 2T_{\pm,i},\ \ i=1,2;\\ \phantom{dsd}[T_{\pm
 1},T_{0,2}]=\pm T_{\pm 1},\ \ [T_{\pm 2},T_{0,1}]=\pm T_{\pm 2};\\
\phantom{ghj}[T_{\pm 1},[T_{\pm 1},T_{\pm 2}]]=
[T_{\pm 2},[T_{\pm 2},T_{\pm 1}]]=0
\ee
and by the commutation relation which defines the third generator
$\pm T_{\pm 12}\equiv [T_{\pm 1},T_{\pm 2}]$
(in order to simplify the notations we sometimes use $\pm T_{\pm 3}$
for this generator).
The first fundamental representation:
$$
T_{0,1}=\left(\begin{array}{ccc}1&0&0\\0&-1&0\\0&0&0\end{array}\right),\
T_{0,2}=\left(\begin{array}{ccc}0&0&0\\0&1&0\\0&0&-1\end{array}\right),
$$
$$
T_{+1}=\left(\begin{array}{ccc}0&1&0\\0&0&0\\0&0&0\end{array}\right),\
T_{+2}=\left(\begin{array}{ccc}0&0&0\\0&0&1\\0&0&0\end{array}\right),\
T_{+12}=\left(\begin{array}{ccc}0&0&1\\0&0&0\\0&0&0\end{array}\right),
$$
\be
T_{-1}=\left(\begin{array}{ccc}0&0&0\\1&0&0\\0&0&0\end{array}\right),\
T_{-2}=\left(\begin{array}{ccc}0&0&0\\0&0&0\\0&1&0\end{array}\right),\
T_{-12}=\left(\begin{array}{ccc}0&0&0\\0&0&0\\1&0&0\end{array}\right).
\ee
Quadratic Casimir operator:
\be
C_2=\sum_{i=1}^3
(T_{-i}T_{+i}+T_{+i}T_{-i})+{2\over 3}
\left(\sum_{i=1}^2T_{0,i}^2+T_{0,1}T_{0,2}
\right)=\\=
2\sum_{i=1}^3T_{-i}T_{+i}+2\sum_{i=1}^2T_{0,i}+
{2\over 3}\left(\sum_{i=1}^2T_{0,i}^2+T_{0,1}T_{0,2}\right).
\ee
In the $SL(3)$ case, we consider only the Gauss
decomposition. Iwasawa decomposition in general $SL(N)$ case is considered
in sect.3.
Then, the parametrization of the group element is
\be
g_G(\{\psi\},\{\phi\},\{\chi\}|T) =\\=
e^{\psi_1 T_{-1}+\psi_2T_{-2}+(\psi_{12}-\psi_1\psi_2)T_{-12}
}e^{\phi_1 T_{0,1}+\phi_2T_{0,2}}e^{\chi_1 T_{+1}+\chi_2T_{+2}+
(\chi_{12}-\chi_1\chi_2)T_{12}}=\\=
e^{\psi_1 T_{-1}}e^{\psi_2T_{-2}}e^{\psi_{12}T_{-12}}
e^{\phi_1 T_{0,1}}e^{\phi_2T_{0,2}}e^{\chi_1 T_{+12}}e^{\chi_2T_{+2}}
e^{\chi_1T_{+1}}.
\ee
Group element in the fundamental representation:
\be\label{groupelGsl3}
g_G = \left( \begin{array}{ccc} 1 & 0 & 0\\ \psi_1 & 1 & 0 \\
\psi_{12} & \psi_2 & 1\end{array}\right)
 \left( \begin{array}{ccc} e^{\phi_1} & 0 & 0 \\
0 & e^{-\phi_1+\phi_2} & 0 \\ 0 & 0 & e^{-\phi_2} \end{array}\right)
   \left( \begin{array}{ccc} 1 & \chi_1 & \chi_{12} \\
   0 & 1 & \chi_2 \\ 0 & 0 & 1\end{array}\right).
\ee
Let us also introduce the invariant notations suitable for any group of rank
$h$ -- simple roots $\balpha_i$, Cartan matrix
$A_{ij}\equiv\balpha_i\cdot\balpha_j$, fundamental weights $\bmu_i$ which are
defined as lying at the dual lattice $\bmu_i\cdot\alpha_j=\delta_{ij}$
(i.e. $\bmu_i= A_{ij}^{-1}\balpha_j$) and
$\brho\equiv \2\sum_{\balpha\in\De^+}\balpha=\sum_i \bmu_i$,
where $\De^+$ is the
subset of the positive roots.
In these notations, $\phi_i=-\bmu_i\cdot\bphi$ and
$C_2= \sum_{{\balpha\in\De^+}}
(T_{-\balpha}T_{+\balpha}+T_{+\balpha}T_{-\balpha})+
\sum_{i=1}^hT_{0,i}A^{-1}_{ij}T_{0,j}$ and $\pm\balpha$
denotes positive and negative roots respectively. For the future needs, we also
introduce the roots $\balpha_i$ as vectors in the $N$-dimensional Cartan plane
of $GL(N)$ group: $\balpha_i=\bfe_{i+1}-\bfe_i$,
$\bfe_i\cdot\bfe_j=\delta_{ij}$.

\subsection{Regular representation}
We consider here only the right regular representation since the left one
is determined just by the interchange $T_{+i}\leftrightarrow T_{-i}$
and $\psi_i\leftrightarrow \chi_i$. The right regular representation is
\be\label{regrep3}
{\cal D}^R_{+1} = \frac{\partial}{\partial \chi_1}, \ \
{\cal D}^R_{+2} = \frac{\partial}{\partial \chi_2} +
         \chi_1 \frac{\partial}{\partial \chi_{12}},\ \
{\cal D}^R_{+12} = \frac{\partial}{\partial \chi_{12}},\\
{\cal D}^R_{0,1} =- 2\chi_1{\d\chi_1}+\chi_2{\d\chi_2}-\chi_{12}{\d\chi_{12}}+
{\d\phi_1},\\
{\cal D}^R_{0,2} = \chi_1{\d\chi_1}-2\chi_2{\d\chi_2}-\chi_{12}{\d\chi_{12}}+
{\d\phi_2},\\
{\cal D}^R_{-1} = e^{\phi_2-2\phi_1}\frac{\partial}{\partial\psi_1} +
e^{\phi_2-2\phi_1}\psi_2\frac{\partial}{\partial\psi_{12}}+\\+
\chi_1{\d\phi_1}-
      \chi_1^2\frac{\partial}{\partial\chi_1} -
      \chi_1\chi_{12}\frac{\partial}{\partial\chi_{12}} -
      (\chi_{12} - \chi_1\chi_2)\frac{\partial}{\partial\chi_2},\\
{\cal D}^R_{-2} = e^{\phi_1-2\phi_2}\frac{\partial}{\partial\psi_2}+
\chi_2{\d \phi_2}+
         \chi_{12}\frac{\partial}{\partial \chi_1} -
          \chi_2^2\frac{\partial}{\partial\chi_2} ,  \nn\\
{\cal D}^R_{-12} = e^{\phi_2-2\phi_1}\chi_2{\d\psi_1}-
e^{\phi_1-2\phi_2}\chi_1{\d\psi_2}
+\left(e^{-\phi_1-\phi_2}+\psi_2\chi_2e^{\phi_2-2\phi_1}\right){\d\psi_{12}}
+\\
+\chi_{12}{\d\phi_1}-(\chi_1\chi_2-\chi_{12}){\d\phi_2} -\chi_1\chi_{12}
{\d\chi_1}+\chi_2(\chi_1\chi_2-\chi_{12}){\d\chi_2}-\chi_{12}^2{\d\chi_{12}}.
\ee
The algebra in the highest weight (irreducible) representation induced by the
one-dimensional representations of the Borel subalgebra
is given by the same formulas (\ref{regrep3})
with all the derivatives ${\d\psi_{*}}$ vanishing and ${\d\phi_i}$ put
equal to $j_i$ ($i=1,2$):
$$
T_{+1} = \frac{\partial}{\partial x_1}, \ \
T_{+2} = \frac{\partial}{\partial x_2} +
         x_1 \frac{\partial}{\partial x_{12}},\ \
T_{+12} = \frac{\partial}{\partial x_{12}},
$$
$$
T_{0,1} =-2x_1{\d x_1}+x_2{\d x_2}-x_{12}{\d x_{12}}+2j_1,
$$
$$
T_{0,2} = x_1{\d x_1}-2x_2{\d x_2}-x_{12}{\d x_{12}}+2j_2,
$$
\be
T_{-1} = 2j_1x_1- x_1^2\frac{\partial}{\partial x_1} -
      x_1x_{12}\frac{\partial}{\partial x_{12}} -
      (x_{12} - x_1x_2)\frac{\partial}{\partial x_2},\\
T_{-2} = 2j_2x_2+
         x_{12}\frac{\partial}{\partial x_1} -
          x_2^2\frac{\partial}{\partial x_2} ,  \nn\\
T_{-12} = 2j_1x_{12}-2j_2(x_1x_2-x_{12})-x_1x_{12}
{\d x_1}+x_2(x_1x_2-x_{12}){\d x_2}-x_{12}^2{\d x_{12}}.
\ee

\subsection{Hamiltonian -- Reducing from regular representation}
Casimir operator is (as given in the regular representation):
\be
\2 C_2={\f 3}\left({\partial^2\over\partial\phi_1^2}+
{\partial^2\over\partial\phi_2^2}+{\partial^2\over\partial\phi_1
\partial\phi_2}\right)+
{\d\phi_1}+{\d\phi_2}+\\+e^{\phi_2-2\phi_1}\left(
{\partial^2\over\partial\psi_1\partial\chi_1}+
\psi_2{\partial^2\over\partial\psi_{12}\partial\chi_1}+
\chi_2{\partial^2\over\partial\psi_1\partial\chi_{12}}+
\psi_2\chi_2{\partial^2\over\partial\psi_{12}\partial\chi_{12}}\right)+\\+
e^{\phi_1-2\phi_2}
{\partial^2\over\partial\psi_{2}\partial\chi_{2}}+
e^{-\phi_1-\phi_2}{\partial^2\over\partial\psi_{12}\partial\chi_{12}}.
\ee
Then, imposing the reduction condition
\be
{\d\chi_{12}}F={\d\psi_{12}}F=0,\ \ {\d\psi_{1,2}}F=i\mu_{1,2}^LF,\ \
{\d\chi_{1,2}}F=i\mu_{1,2}^RF,
\ee
i.e.
\be\label{statecondsl3}
T_{+1,2}|\psi_R>=i\mu^R_{1,2}|\psi_R>,\
T_{+12}|\psi_R>=0,\\  <\psi_L|T_{-1,2}=i\mu^L_{1,2}<\psi_L|,\
<\psi_L|T_{-12}=0,
\ee
we obtain the Hamiltonian
\be
\2 H={\f 3}\left({\partial^2\over\partial\phi_1^2}+
{\partial^2\over\partial\phi_2^2}+{\partial^2\over\partial\phi_1\partial\phi_2}
\right)+{\d\phi_1}+{\d\phi_2}-\\-
\mu_1^L\mu_1^Re^{\phi_2-2\phi_1}-\mu_2^L\mu_2^Re^{\phi_1-2\phi_2}.
\ee
Then, the scaled matrix element $\Psi(\phi_1,\phi_2)\equiv e^{\phi_1+\phi_2}F$
satisfies the Liouville equation
\be\label{SchGsl3}
\left({\f 3}\left({\partial^2\over\partial\phi_1^2}+
{\partial^2\over\partial\phi_2^2}+{\partial^2\over\partial\phi_1
\partial\phi_2}\right)-\mu_1^L\mu_1^R
e^{\phi_2-2\phi_1}-\mu_2^L\mu_2^R
e^{\phi_1-2\phi_2}\right)\Psi(\phi_1,\phi_2)=\\=
\left(2(j_1+j_2)+{4\over 3}(j_1^2+j_2^2+j_1j_2)+1\right)\Psi(\phi_1,\phi_2).
\ee
Being invariantly written, this equation takes the form
\be
\left(A^{-1}_{ij}{\partial^2\over\partial\phi_i\partial\phi_j}
-2\sum_i\mu_i^L\mu_i^Re^{\balpha_i\cdot\bphi}\right)\Psi(\phi_1,\phi_2)=\\=
(2\bj+\brho)^2\Psi(\phi_1,\phi_2)\equiv
\blambda^2\Psi(\phi_1,\phi_2),
\ee
where $\Psi(\phi_1,\phi_2)\equiv e^{-\brho\cdot\bphi}F$, $\bj\equiv
j_i\bmu_i$ and the repeated indices are summed over (eigenvalue of the
quadratic Casimir operator in these notations is $2\bj(\bj+\brho)$).

In the Iwasawa case, one can easily obtain the Liouville equation which differs
from (\ref{SchGsl3}) by additional factor 2 in the exponents. This means that
its solution can be obtained from the Gauss one by the replace $2\mu_i^L\mu_i^R
\to \mu_i^2$ and $2\bj+\brho\to (\bj+{\brho\over 2})$, i.e. $2j_i+1\to j_i+\2$.
It is very important that, under this replace, pure imaginary combination
remains pure imaginary.

\subsection{Liouville wave function (LWF) -- Gauss Whittaker LWF}
In order to avoid the repetition of clear but tedious calculations, we
consider here only Gauss LWF. The Iwasawa $SL(3)$ LWF can be easily obtained
from the results for general $SL(N)$ group below.

As before, we can rewrite reduction conditions (\ref{statecondsl3})
as given on the states in the highest weight representation. Then, we
obtain the equations
\be
{\d x_1}\psi_R(x_1,x_2)= i\mu_1^R\psi_R(x_1,x_2),\ \
{\d x_2}\psi_R(x_1,x_2)= i\mu_2^R\psi_R(x_1,x_2)
\ee
and
\be
\left(x_1^2{\d x_1}-(x_1x_2-x_{12}){\d x_2}+x_1x_{12}{\d x_{12}}+
2(j_1+1)\right)\psi_L(x_1,x_2,x_{12})=\\=
i\mu_1^L\psi_L(x_1,x_2,x_{12}),\\
\left(-x_{12}{\d x_1}+x_2^2{\d x_2}+2(j_2+1)x_2\right)\psi_L(x_1,x_2,x_{12})=
i\mu_2^L\psi_L(x_1,x_2,x_{12}),\\
\left(x_1x_{12}{\d x_1}-x_2(x_1x_2-x_{12}){\d x_2}+x_{12}^2{\d x_{12}}+\right.
\\ \left.\phantom{{\d x_1}}+
2(j_1+j_2+2)x_{12}-2(j_2+1)x_1x_2\right)\psi_L(x_1,x_2,x_{12})=0.
\ee
These equations have the following solution:
\be
\psi_R(x_1,x_2)=e^{i\mu_1^Rx_1+i\mu_2^Rx_2},\\
\psi_L(x_1,x_2,x_{12})=(x_{12}-x_1x_2)^{-2(j_2+1)}x_{12}^{-2(j_1+1)}
e^{-i\mu_2^L{x_1\over x_{12}}-i\mu_1^L{x_2\over x_1x_2-x_{12}}}.
\ee
Similarly to the $SL(2)$ case, the measure in $x$-variables is flat, and
the integration contour is the real semi-axis. Then, we finally get
the LWF as the matrix element
$$
\Psi(\phi_1,\phi_2)=e^{\phi_1+\phi_2}\int \ dx_1dx_2dx_{12}\
(x_{12}-x_1x_2)^{-2(j_2+1)}x_{12}^{-2(j_1+1)}\times
$$
$$
\times
e^{-i\mu_2^L{x_1\over x_{12}}-i\mu_1^L{x_2\over x_1x_2-x_{12}}}
\exp\left\{\phi_1\left(-2x_1{\d x_1}+x_2{\d x_2}-x_{12}{\d x_{12}}+2j_1\right)+
\right.
$$
$$
+\left.\phi_2\left(x_1{\d x_1}-2x_2{\d x_2}-x_{12}{\d x_{12}}+2j_2
\right)\right\}e^{i\mu_1^Rx_1+i\mu_2^Rx_2}=
$$
\be\label{psisl3}
=e^{(2j_1+1)\phi_1+(2j_2+1)\phi_2}\int \ dx_1dx_2dx_{12}\
(x_{12}-x_1x_2)^{-2(j_2+1)}x_{12}^{-2(j_1+1)}\times\\\times
e^{-\mu_2^L{x_1\over x_{12}}-\mu_1^L{x_2\over x_1x_2-x_{12}}-
\mu_1^Rx_1e^{\phi_2-2\phi_1}-\mu_2^Rx_2e^{\phi_1-2\phi_2}}.
\ee
After the change of variables $x_{12} = x_1x_2t$, this integral
acquires a more invariant form:
\be\label{psisl3'}
e^{(2j_1+1)\phi_1+(2j_2+1)\phi_2}
\int dx_1dx_2dt (x_1x_2)^{-2(j_1+j_2+2)+1}
(1-t)^{-2(j_2+1)}t^{-2(j_1+1)}\times\\\times
e^{-{\mu_2^L\over tx_2}-{\mu_1^L\over (1-t)x_1}-
\mu_1^Rx_1e^{\phi_2-2\phi_1}-\mu_2^Rx_2e^{\phi_1-2\phi_2}}\sim\\\sim
e^{(j_1-j_2)\phi_1+(j_2-j_1)\phi_2}\int dt\ (1-t)^{-2(j_2+1)}t^{-2(j_1+1)}
\times\\\times
K_{2(j_1+j_2+1)}\left(2\mu_1^L\mu_1^R{e^{{\phi_2\over 2}-\phi_1}\over
\sqrt{1-t}}\right)
K_{2(j_1+j_2+1)}\left(2\mu_2^L\mu_2^R{e^{{\phi_1\over
2}-\phi_2}\over \sqrt{t}}\right),
\ee
where we used formula (\ref{AMac}).

\subsection{Asymptotics and Harish-Chandra functions}
In order to find Harish-Chandra functions, we need to calculate asymptotics of
LWF. In the $SL(3)$ case there are 6 different asymptotics, their number being
given by the number of elements of the corresponding Weyl group. Let us see
how it works. We should consider LWF (\ref{psisl3}) in the asymptotical region
where potential is zero, i.e. both exponentials vanish. Still the calculation
depends on the signs of $\lambda_i\equiv 2j_i+1$. The simplest case is when
both of them are positive. Then, one can easily get (using formulas of
Appendix A)

\be\label{assl3}
\Psi(\phi_1,\phi_2)\to {\pi^3\over \sin\lambda_1\sin\lambda_2
\sin(\lambda_1+\lambda_2)}
\left(\mu_1^L\mu_2^L\right)^{-2(\lambda_1+\lambda_2)}\times\\
\times{\f \Gamma(1-\lambda_1)
\Gamma(1-\lambda_2)\Gamma(1-\lambda_1-\lambda_2)}e^{\lambda_1\phi_1+
\lambda_2\phi_2}.
\ee
Now let us consider other possible values
of $\lambda_i$. Totally, there are 6 essentially different domains:

\be
\lambda_1>0, \ \ \ \lambda_2>0;\\
\lambda_1>0, \ \ \ \lambda_2<0, \ \ \ \lambda_1+\lambda_2>0;\\
\lambda_1>0, \ \ \ \lambda_2<0, \ \ \ \lambda_1+\lambda_2<0;\\
\lambda_1<0, \ \ \ \lambda_2>0, \ \ \ \lambda_1+\lambda_2>0;\\
\lambda_1<0, \ \ \ \lambda_2>0, \ \ \ \lambda_1+\lambda_2<0;\\
\lambda_1<0, \ \ \ \lambda_2<0.
\ee
In each of these domains one should calculate asymptotics of the integral
(\ref{psisl3}) redefining properly variables like it has been done in s.1.4.3
for $SL(2)$ case. It is evident that each asymptotics is described by the
$\lambda$'s lying in the corresponding Weyl chamber, and, therefore, the number
of asymptotics is equal to the number of elements of the Weyl group.
It is natural to introduce the third $\lambda_3\equiv \lambda_1+\lambda_2$.
Now, choosing properly the normalization of LWF to cancel all the poles
(see s.1.4.6), we
finally obtain 6 different Harish-Chandra functions:

\be\label{hchsl3}
c_s=\prod_i{\f \Gamma(1-s\lambda_i)},
\ee
where $s$ means Weyl group element and the product goes over all the positive
roots (i.e. $\lambda_1$, $\lambda_2$ and $\lambda_3$).

\subsection{LWF by the Fourier transform}
To complete this section, we consider the solution to the $SL(3)$ Liouville
equation in momentum representation. Let us first introduce instead of
$\phi_i$ new variables $\xi_i\equiv \balpha_i\cdot\phi$. Now,
after making the Fourier
transform with respect to these variables (i.e. $f(\xi_1,\xi_2) =
\int dp_1dp_2 e^{ip_1\xi_1+ip_2\xi_2}\bar f(p_1,p_2)$) and putting $\mu_i^L
\mu_i^R=1$, the Liouville equation takes the following
form\footnote{Note the different signs in the bilinears of $p$ and
$\lambda$. This is the point which leads to the non-trivial
solution to the Fourier Liouville equation.}:

\be
-\left(p_1^2+p_2^2-p_1p_2+{\f 3}(\lambda_1^2+
\lambda_2^2+\lambda_1\lambda_2)\right)\bar f(p_1,p_2)=\\
=\bar f(p_1+i,p_2)+
\bar f(p_1,p_2+i),
\ee
where we used that $(\bj+\brho)^2={1\over 3}(\lambda_1^2+\lambda^2_2+
\lambda_1\lambda_2)$. Let us first solve this equation at $\lambda_1=
\lambda_2=0$. The solution is

\be
\bar f(p_1,p_2)={\Gamma^3\left({p_1\over i}\right)
\Gamma^3\left({p_2\over i}\right)\over \Gamma\left({p_1+p_2\over i}\right)}=
\Gamma^2\left({p_1\over i}\right)B\left({p_1\over i},{p_2\over i}\right)
\Gamma^2\left({p_2\over i}\right).
\ee
Now it is already easy to find solutions at any $\lambda_i$ fitting proper
shifts of the $\Gamma$-function arguments (let us point out
again that in order to get the general solution, one needs to multiply this one
by an arbitrary periodic function). The result reads
$$
\bar f(p_1,p_2)=
\left[\Gamma\left({p_1+p_2\over i}\right)\right]^{-1}
\Gamma\left({p_1\over i}-\left({2\over 3}
\lambda_1+{\f 3}\lambda_2\right)\right)\times
$$
$$
\times
\Gamma\left({p_1\over i}+
\left({2\over 3}
\lambda_2+{\f 3}\lambda_1\right)\right)
\Gamma\left({p_1\over i}+
\left({1\over 3}
\lambda_1-{\f 3}\lambda_2\right)\right)
\times
$$
$$
\times
\Gamma\left({p_2\over i}+\left({2\over 3}
\lambda_1+{\f 3}\lambda_2\right)\right)
\times
$$
$$
\times
\Gamma\left({p_2\over i}-
\left({2\over 3}
\lambda_2+{\f 3}\lambda_1\right)\right)
\Gamma\left({p_2\over i}-
\left({1\over 3}
\lambda_1-{\f 3}\lambda_2\right)\right)=
$$
\be\label{fsl3}
= \left[\Gamma\left({p_1+p_2\over i}\right)\right]^{-1}
\Gamma\left({p_1\over i}-\bar\lambda_1\right)
\Gamma\left({p_1\over i}+\bar\lambda_2\right)
\Gamma\left({p_1\over i}+\bar\lambda_1-\bar\lambda_2\right)
\times\\\times
\Gamma\left({p_2\over i}+\bar\lambda_1\right)
\Gamma\left({p_2\over i}-\bar\lambda_2\right)
\Gamma\left({p_2\over i}-\bar\lambda_1+\bar\lambda_2\right)=\\=
\Gamma\left({p_1\over i}-\bar\lambda_1\right)
\Gamma\left({p_1\over i}+\bar\lambda_2\right)
B\left({p_1\over i}+\bar\lambda_1-\bar\lambda_2,
{p_2\over i}-\bar\lambda_1+\bar\lambda_2\right)\times
\\\times\Gamma\left({p_2\over i}+\bar\lambda_1\right)
\Gamma\left({p_2\over i}-\bar\lambda_2\right),
\ee
where $\bar\lambda_1\equiv {2\over 3}\lambda_1+{1\over 3}\lambda_2$,
$\bar\lambda_2\equiv {2\over 3}\lambda_2+{1\over 3}\lambda_1$ are the
projections of the vector $\blambda\equiv\lambda_i\bmu_i$ onto the root
vectors,
i.e. $\bar\lambda_i=A_{ij}^{-1}\lambda_j$. The result (\ref{fsl3}) looks as
a product over the projections to all 6 vectors.
Again one can trivially obtain the asymptotics of this solution
looking at the poles of the $\Gamma$-functions. The result evidently coincides
with (\ref{assl3}). One can also immediately demonstrate like it has been done
for the $SL(2)$ case that the inverse Fourier transform of this solution leads
to (\ref{psisl3}). Indeed, one should use the last two lines of (\ref{fsl3}),
the second integral representation in (\ref{AGamma}) for the $B$-function
and the integral representation for $\Gamma$-functions to prove that

\be\label{sl3fg}
f(\bxi)=\int\ldots\int dx_1d\tilde x_1dx_2d\tilde x_2dp_1dp_2dt
x_1^{{p_1\over i}-\bar\lambda_1-1}\tilde x_1^{{p_1\over i}+\bar\lambda_2-1}
x_2^{{p_2\over i}-\bar\lambda_2-1}\tilde \times\\\times
x_2^{{p_2\over i}+\bar\lambda_1-1}
(1-t)^{{p_1\over i}+\bar\lambda_1-\bar\lambda_2-1}t^{{p_2\over i}-\bar\lambda_1
+\bar\lambda_2-1}
e^{-x_1-x_2-\tilde x_1-\tilde x_2}e^{ip_1\xi_1+ip_2\xi_2}=\\=
\int\ldots\int {dx_1\over x_1} {d\tilde x_1\over \tilde x_1}
{dx_2\over x_2}{d\tilde x_2\over \tilde x_2} {dt\over t(1-t)}dp_1dp_2
\left(x_1\tilde x_1(1-t)e^{-\xi_1}\right)^{{p_1\over i}}
\times\\\times
\left(x_2\tilde x_2 te^{-\xi_2}\right)^{{p_2\over i}}
\left({\tilde x_2(1-t)\over x_1 t}\right)^{\bar\lambda_1}
\left({\tilde x_1 t\over x_2(1-t)}\right)^{\bar\lambda_2}
e^{-x_1-\tilde x_1-x_2-\tilde x_2}=\\=
\int\int\int {dx_1\over x_1}{dx_2\over x_2}{dt\over t(1-t)}
\left({(1-t)e^{\xi_2}\over x_1x_2t^2}\right)^{\bar\lambda_1}
\left({t e^{\xi_1}\over x_1x_2(1-t)^2}\right)^{\bar\lambda_2}
\times\\\times
e^{-x_1-x_2-{e^{\xi_1}\over x_1(1-t)}-{e^{\xi_2}\over x_2t}},
\ee
which coincides with (\ref{psisl3'}).

\section{$SL(N)$}
\setcounter{equation}{0}
\subsection{Notations}
Algebra is completely given by the non-zero simple root commutation
relations:
\be\label{commrel1}
\phantom{fhg}[T_{\pm i},T_{0,j}]=\mp A_{ij}T_{\pm,i},\ \ [T_{+i},T_{-j}]=
\delta_{ij}T_{0,j},\ \ i,j=1,\ldots, N-1,
\ee
and the Serre relations
\be\label{Serre}
\phantom{fhg}\hbox{ad}_{T_{\pm i}}^{1-A_{ij}}\left(T_{\pm j}\right)=0,
\ee
where $\hbox{ad}_x^k(y)\equiv \underbrace{[x,[x,...,[x,y]..]]}_{k\ \hbox{
times}}$.
All other commutation relations can be obtained from
(\ref{commrel1})-(\ref{Serre}),
the generators which corresponds to positive (negative) non-simple roots
being constructed from the positive (negative) simple root generators
by the manifest formula
$[T_{\balpha},T_{\bbeta}]=N_{\balpha,\bbeta}T_{\balpha+\bbeta}$.
Here the generator $T_{\balpha+\bbeta}$ corresponds to the
non-simple root $\balpha+\bbeta$ and $N_{\balpha,\bbeta}$ are some
non-zero structure constants. With using
the non-simple root generators, the Serre identities are replaced
by appropriate Lie algebra relations.

Quadratic Casimir operator is
\be
C_2=\sum_{\balpha\in\De}
T_{\balpha}T_{-\balpha}+\sum_{ij}^{N-1}A_{ij}^{-1}T_{0,i}T_{0,j},
\ee
where the first sum goes over all (positive and negative) roots.

\subsection{Representations}
In the case of generic $SL(N)$ group, one can define the
(right) regular representation only in general terms of the group acting on the
space of the algebra of functions:
\beq\label{rrrep}
\pi_{reg}(h) f(g)=f(gh).
\eeq
Therefore, we use from now on mostly group (not algebra) terms. Still, we can
restrict the space of functions to the irreducible representations in the
generic situation. For doing this, we consider the representation induced by
one-dimensional representations of the Borel subgroup. That is, we reduce the
space of all functions to the functions satisfying the following covariance
property:
\beq\label{irrep}
f_{\lambda}(bg)=\chi _{\lambda}(b)f_{\lambda}(g),
\eeq
where $b$ is an element of
the Borel subgroup of lower-triangle matrices and $\chi _{\lambda}$ is the
character of the Borel subgroup of the form:
\beq \label{rep}
\chi
_{\lambda}(b)=\prod _{i=1}^{N-1} \mid b_{ii} \mid ^{(\blambda-\brho)\bfe_i}
(\mbox{sign}b_{ii} )^{\epsilon _i},
\eeq
where $\epsilon_i$ are equal to either 0 or 1. For the sake of simplicity, we
consider the representations with all these sign factors to be zero although
other cases can be also easily treated. The representation constructed belongs
to the principal (spherical) series.

Thus, our representation is given by
restricting the space of functions to the functions defined on the coset
$B\backslash G$ which, in turn, may be identified with the strictly
upper-triangular matrices $N_+$. At given $\lambda$, there is a natural
Hermitian bilinear form on the space of matrix elements of $X$ which is given
just by the flat measure:

\be
<f_L|f_R>_{\lambda}
=\int_{X=B\backslash G} \overline {f_{L,\lambda}(x)}f_{R,\lambda}(x)
\prod_{ij}dx_{ij}.
\ee
This form becomes a scalar product provided by the pure imaginary
$\lambda$'s.
This gives us unitary irreducible representations of the main (spherical)
series.

Now let us describe the structure of the fundamental representations of
$SL(N)$, some formulas being used in sect.3.8.

Let us consider the upper-triangle $N\times N$-matrix with the unit diagonal,
\be
||X||_{ij} = x_{ij}\theta(j-i)\;\;, \;\;\;\;\;x_{ii}\equiv 1\; .
\ee
Let also
\be
\Delta_{i_1,\ldots ,i_k}(X) \equiv \left|
\begin{array}{lll}
x_{1,i_1} & \ldots &x_{1,i_k}\\
\ldots    &  \ldots  &  \ldots\\
x_{k,i_1} & \ldots &x_{k,i_k}
\end{array}\right|  \;\;\;\; 
\ee
These minors are considered as functions on $SL(N)$. Obviously, they
are left invariant with respect to the action of the nilpotent subgroup
$N_{-}\subset SL(N)$:
\be
\Delta_{i_1,\ldots ,i_k}(g_{-}X)  = \Delta_{i_1,\ldots ,i_k}(X),
\ee
where $g_{-}\in N_{-}$. Further, the left multiplication by the diagonal
matrix $D = {\rm diag}(d_1, \ldots , d_N)\in SL(N)$ acts as
\be\label{bor}
\Delta_{i_1,\ldots ,i_k}(DX)  = \prod_{j=1}^{k}d_j\;
\Delta_{i_1,\ldots ,i_k}(X)\;.
\ee
In other words, this implies that
\be
\Delta_{i_1,\ldots ,i_k}(e^{\sum t_jT_{0,j}}X)  = e^{t_k}\;
\Delta_{i_1,\ldots ,i_k}(X)\;.
\ee
Thus, in accordance with (\ref{irrep}), (\ref{rep}),
all the minors $\;\Delta_{i_1,\ldots ,i_k}(X)\;$ with fixed $\;k\;$
belong to the space of the $k$-th fundamental representation, ${\cal F}_k$.
As a particular case, one can consider the following minors:
\be
\Delta_k(X) \equiv \left|
\begin{array}{lll}
x_{1,N-k+1} & \ldots &x_{1N}\\
\ldots    &  \ldots  &  \ldots\\
x_{k,N-k+1} & \ldots &x_{kN}
\end{array}\right|  \;,\;\;\; k = 1, \ldots , N\;\; .
\ee
These minors are right invariant with respect to the action of $N_{-}$:
\be\label{inv}
\Delta_k(Xe^{tT_{-,i}}) = \Delta_k(X) \; ,
\ee
where $T_{-,i}$ are lowering generators of $SL(N)$ associated with the
simple roots; therefore,
$\Delta_k \in {\cal F}_k$ is the lowest weight vector of the
$k$-th fundamental representation. Obviously, the weight of
$\Delta_k$ is determined from the formula
\be\label{weight}
\Delta_k(Xe^{\sum t_jT_{0,j}})  = e^{-t_{N-k}}\;
\Delta_k(X)\;.
\ee

\subsection{Hamiltonians and Liouville equation}
In this subsection we derive Liouville equation for $SL(N)$ by the
immediate matrix element calculation (cf. 1.3.2). Namely, we impose the
following reduction conditions:

\bigskip

\noi
In the Iwasawa case:

\be\label{rcglNIR}
T_{+,i}|\psi_R>=i\mu_i|\psi_R>
\ee
and
\be\label{rcglNIL}
<\psi_L|T_{+,i}=<\psi_L|T_{-,i}, \ \ \ \hbox{i.e.}\ \ \
<\psi_L|k=0,
\ee
where $k$ denotes any algebraic element of the maximal compact subgroup $K$
of $SL(N)$.

\bigskip

\noi
In the Gauss case:

\be\label{rcglNGR}
T_{+,i}|\psi_R>=i\mu_i^R|\psi_R>
\ee
and
\be\label{rcglNGL}
<\psi_L|T_{-,i}=i\mu_i^L<\psi_L|.
\ee
These constraints are given only by the simple root generators, since
all the rest are generated by the commutation relations. Say, the action
of non-simple roots generators cancels both the left and right states
in the Gauss case and the right state in the Iwasawa case etc.

Now, in the Iwasawa case, one obtains

\be
(\blambda^2-\brho^2)F^{(\blambda)}_I(\bphi_I)\equiv
(\blambda^2-\brho^2)<\psi_L|e^{-\bmu_i\bphi_IT_{0,i}}|\psi_R>=\\=
<\psi_L|e^{-\bmu_i\bphi_IT_{0,i}}C_2|\psi_R>=
<\psi_L|e^{-\bmu_i\bphi_IT_{0,i}}\left(2\sum_{\balpha\in\De^+}
T_{\balpha}T_{-\balpha}+ \right.\\\left.+2\sum_{ij} A^{-1}_{ij}T_{0,j}
+\sum_{ij}A_{ij}^{-1}T_{0,i}T_{0,j}\right)|\psi_R>=\\=
\left( {\partial^2\over\partial\bphi^2_I}+ 2\sum_i{\d (\balpha_i\bphi_I)}
-2\sum_i\mu^2_ie^{2\balpha_i\bphi_I}
\right)<\psi_L|e^{-\bmu_i\bphi_IT_{0,i}}|\psi_R>.
\ee
Therefore, the combination
$\Psi^{(\blambda)}_I(\bphi_I) e^{-\brho\bphi_I}F^{(\blambda)}_I(\bphi_I)$
satisfies the following Liouville equation
\be\label{LeslNI}
\left( {\partial^2\over\partial\bphi^2_I}
-2\sum_i\mu^2_ie^{2\balpha_i\bphi_I}
\right)\Psi^{(\blambda)}_I(\bphi_I)=\blambda^2
\Psi^{(\blambda)}_I(\bphi_I).
\ee
Similarly, in the Gauss case, one gets
\be
(\blambda^2-\brho^2)F^{(\blambda)}_G(\bphi)\equiv
(\blambda^2-\brho^2)<\psi_L|e^{-\bmu_i\bphi T_{0,i}}|\psi_R>=
<\psi_L|e^{-\bmu_i\bphi T_{0,i}}C_2|\psi_R>=\\=
<\psi_L|e^{-\bmu_i\bphi T_{0,i}}\left(2\sum_{\balpha\in\De^+}
T_{\balpha}T_{-\balpha}+ 2\sum_{ij} A^{-1}_{ij}T_{0,j}
+\sum_{ij}A_{ij}^{-1}T_{0,i}T_{0,j}\right)|\psi_R>=\\=
\left( {\partial^2\over\partial\bphi^2}+ 2\sum_i{\d (\balpha_i\bphi)}
-2\sum_i\mu^L_i\mu_i^Re^{\balpha_i\bphi}
\right)<\psi_L|e^{-\bmu_i\bphi T_{0,i}}|\psi_R>.
\ee
and $\Psi^{(\blambda)}_G(\bphi)=e^{-\brho\bphi}F^{(\blambda)}_G(\bphi)$
satisfies the Liouville equation
\be\label{LeslNG}
\left( {\partial^2\over\partial\bphi^2}
-2\sum_i\mu^L_i\mu_i^Re^{\balpha_i\bphi}
\right)\Psi^{(\blambda)}_G(\bphi)=\blambda^2
\Psi^{(\blambda)}_G(\bphi).
\ee

\subsection{On different solutions to the Liouville equations}
Throughout this paper we discuss the solutions to the Liouville equation
which arise within the group theory framework. Each equation has only one
solution of such a type, while generally there is a lot of solutions to
the equation. In this subsection, we are going to discuss what conditions
select out this unique solution.

Let us start from the simplest $SL(2)$ case. Then, the Liouville equation
(\ref{Le}) has two linearly independent solutions. One can choose, say,
$K_\lambda(e^\varphi)$ and $I_\lambda(e^\varphi)$. From the viewpoint of
integral representations we use in the paper, this means different choices of
integration contours. One can choose two linearly independent
(closed\footnote{In the compactified complex plane.}) contours. However,
requiring the solution to be restricted function over the real axis (the
standard quantum mechanics boundary conditions), one is left with the only
solution $K_\lambda(e^\varphi)$. This choice just corresponds to the group
theory calculation, because of unitarity of the representations we consider.

As an illustrative example, let us see how the two linearly independent
solutions arises in the course of solving the Liouville equation by the
Fourier transform. The general solution is solution (\ref{sl2F}) multiplied
by an arbitrary periodic function $\Phi(p)$:
\be
\bar f(p)=\Gamma ({p\over 2i}+\lambda)\Gamma ({p\over 2i}-\lambda)\Phi(p).
\ee
Such a function can be expanded into the Fourier series:
\be
\Phi(p)=\sum_n c_ne^{\pi np}.
\ee
Now we repeat the procedure of sect.1.4.4. It is easily to see that the
result is
\be
f(\varphi)=\int dp e^{ip\varphi} \Gamma
({p\over 2i}+\lambda)\Gamma ({p\over 2i}-\lambda)\Phi(p)=\\=
\sum_nc_n\int_0^{\infty}dx\int_0^{\infty}d\tilde x\int_{-\infty}^{\infty}dp
e^{-x-\tilde x}
x^{{p\over 2i}-\lambda-1}\tilde x^{{p\over 2i}+\lambda-1}e^{ip\varphi+
\pi pn}
\sim\\\sim
\left(\sum_n c_{2n}\right)K_{2\lambda}(2e^{\phi})+
\left(\sum_n c_{2n+1}\right)K_{2\lambda}(-2e^{\phi}).
\ee
Since $K_{2\lambda}(-z)=e^{-2\pi\lambda i}K_{2\lambda}(z)-
i\pi I_{2\lambda}(z)$, the result is the arbitrary linear combination of the
two
independent solutions to the Liouville equation. That is quite amazing that
the restricted solution is the simplest one, that with $\Phi(p)=1$.

The more serious freedom in choosing the solutions to the Liouville
equation arises for the higher rank groups $SL(N)$. In this case, the space of
solutions is parametrized by $N-2$ arbitrary parameters. In order to
understand what choice of these parameters corresponds to the group theory
approach, one needs to observe that the LWF obtained from the group
theory satisfies also the $N-2$ additional equations generated by the higher
Casimir operators (totally there are right $N-2$ independent higher Casimir
operators). These additional equations, along with the boundary conditions
(restricted solutions), uniquely fix the solution.

\subsection{Equivalence of the Iwasawa and Gauss Whittaker functions}
Now one can use the higher Casimir equations to prove that
the Iwasawa and Gauss Whittaker
functions always coincide, i.e. they are nothing but different integral
representations for the same function. Indeed, one needs only to prove that
they satisfy the same Liouville (quadratic Casimir) and higher Casimir
equations.  Let us illustrate how it works for the first two equations.

In fact, as for the first equation -- Liouville equation itself -- it is
evident from comparing (\ref{LeslNI}) and (\ref{LeslNG}) that they are related
by the replace
\be\label{replace}
2\phi_I\to\phi,\ \ \ \mu_i^2\to 2\mu_i^L\mu_i^R,\ \ \ \blambda\to
2\blambda.
\ee
This which proves the equivalence of the corresponding Whittaker
functions in the $SL(2)$ case. To deal with higher equations, we need to
define the higher Casimir operators for $SL(N)$ group.
For doing this, let us fix some representation $\rho$ of
the group (in fact, one needs to fix some representation of the universal
enveloping algebra). Then, define $L$-operator \cite{FRT}

\be
L\equiv \sum_{\balpha\in\De}\rho(T_{\balpha})\otimes T_{-\balpha}+
\sum_iA^{-1}_{ij}\rho(T_i)\otimes T_j.
\ee
Now, the $k$-th Casimir operator can be defined as

\be
C_k\equiv \Tr_\rho L^k,
\ee
where trace is taken over the representation $\rho$. Since the result does not
depend on the choice of the representation $\rho$, one can take the simplest
one. Let us look at the first fundamental representation and $SL(3)$ group and
calculate the third Casimir operator. For the sake of space, we do not write
down the complete answer, however, the result for the Schr\"odinger equation
obtained by the procedure of s.1.3.2 is

\be
\left({\partial^3\over\partial^2\xi_1\partial\xi_2}-
{\partial^3\over\partial\xi_2^2\partial\xi_1}+2{\partial^2\over\partial\xi_1^2}
-{\partial^2\over\partial\xi_1\partial\xi_2}+\right.\\\left.+2{\d
\xi_1}+2\eta_1+ \eta_1{\d \xi_2} - \eta_2{\d
\xi_1}\right)F^{(\blambda)}(\bphi)=\\=
\left({\lambda_1^3-\lambda_2^3+(\lambda_1-\lambda_2)^3\over 27}
+{1\over 3}(\lambda_1^2+\lambda_2^2+\lambda_1\lambda_2)-1\right)
F^{(\blambda)}(\bphi),
\ee
where $\xi_i\equiv\balpha_i\bphi$ and $\eta_i\equiv -\mu^2_ie^{2\xi_i}$ for the
Iwasawa case and $\eta_i\equiv -\mu_i^L\mu_i^Re^{\xi_i}$ for the Gauss one.
Using this equation, one obtains the following equation for
$\Psi^{(\blambda)}(\bphi)=e^{-\brho\bphi}F^{(\blambda)}(\bphi)$

\be\label{Le3C}
\left({\partial^3\over\partial^2\xi_1\partial\xi_2}-
{\partial^3\over\partial\xi_2^2\partial\xi_1}+
\eta_1{\d \xi_2} - \eta_2{\d \xi_1}\right)\Psi^{(\blambda)}(\bphi)=\\
={\lambda_1^3-\lambda_2^3+(\lambda_1-\lambda_2)^3\over 27}
\Psi^{(\blambda)}(\bphi),
\ee
where we used Liouville equations (\ref{LeslNI}) and (\ref{LeslNG}).
Equation (\ref{Le3C}) is the only additional equation arising in the
$SL(3)$ case to fix the solution to the Liouville equation (one arbitrary
function), and this equation for the Gauss case is again obtained from the
Iwasawa case equation by the same replace (\ref{replace}). This proves the
equivalence of the Iwasawa and Gauss Whittaker functions in the $SL(3)$ case.

\subsection{Iwasawa Whittaker LWF}
In order to get LWF in the Iwasawa case, as above one needs to construct the
solution to the conditions which describes two functions $f_{L,\lambda}(g)$ and
$f_{R,\lambda}(g)$ defining left and right states respectively:

\be\label{statecondRN}
\pi_{\lambda}(z) f_{R,\lambda}(g)=f_{R,\lambda}(gz)=e^{\sum
i\mu_iz_{i,i+1}}f_{R,\lambda}(g) =e^{iTr (\mu z)}f_{R,\lambda}(g),\\ (\mu
)_{ij}=\delta _{i-1,j}\mu _i, \ \ \ g,z\in N_+,
\ee
and
\be\label{statecondLN}
\pi_{\lambda}(k)
f_{L,\lambda}(g)=f_{L,\lambda}(g),\ \ \ k\in K,\ \ \ g\in N_+,
\ee
the algebraic
versions of these relations being (\ref{rcglNGL}) and (\ref{rcglNIR})
(in the first relation
we used formula (\ref{rrrep})).  To solve the first condition is the same as to
find a one-dimensional representation of the group of upper-triangular
matrices. First, we construct the
additive character of this group using the fact that in the product of two
upper-triangular matrices, their elements next to the main
diagonal are summed. Then, we exponentiate this character to get finally the
one-dimensional representation:
\beq\label{fR}
f_{R,\lambda}^{\mu}(x)=e^{i\Tr \mu x}.
\eeq
The second condition can be solved along the following line. Let us consider
the Iwasawa expansion of an element $x\in N_+$:

\be\label{expansion}
x=n_-\cdot h\cdot k,
\ee
where $n_-$ is an element of the maximal nilpotent subgroup $N_-$, $h$ is an
element of the Cartan subgroup $H$ and $k$ is an element of the maximal
compact subgroup $K$.
Then,

\be
f_{L,\lambda}(x)=
f_{L,\lambda}(n_- h k)= \chi _{\lambda}(n_- h) f_{L,\lambda}(k)=\\=
\chi _{\lambda}(n_- h)=\chi _{\lambda}(h)=h_i^{\lambda_i-1}\equiv
h^{(\blambda-\brho)\bfe_i}.
\ee
The first equality follows from (\ref{expansion}), and the second one does from
(\ref{statecondLN}). Now we need to express $h_i$ through matrix elements of
$x$. Indeed, let us make use of (\ref{expansion}) and obtain the expression for
the symmetric matrix $xx^t$, where index $t$ means transponed matrix:
\be
xx^t=n_+h^2n_+^t.
\ee
Now, denoting $\Delta_i(xx^t)$ the upper principal minors of the matrix
$xx^t$, one can easily check that\footnote{To derive this expression, one needs
note that the diagonal element $g_{ii}$ of the matrix $g$
depends only on the matrix elements lying in the
$i\times i$ submatrix in the left upper corner. Then, one may use the induction
over the rang of the matrix to obtain formula (\ref{fL}).}

\be\label{fL}
h^2_i={\Delta_i\over\Delta_{i-1}},\ \ \ h^2_1=\Delta_1,\ \ \ \Delta_N=
{\De_1\over\De_{N-1}}.
\ee
Thus, we have manifestly calculated $f_R(x)$ ((\ref{fR})) and $f_L(x)$
((\ref{fL})). In order to obtain LWF, we now only need to fix the action of
the Cartan part of the group element $g$ on $f_R(x)$. Let us note that,
although the element $xh$ with $x\in X$ and $h\in H$ does not belong to $X$,
the element $h^{-1}xh$ does. Therefore, using (\ref{rrrep}), one can get

\be\label{Cartan}
\pi_{\lambda}(h)f_{R,\lambda}(x)=
f_{R,\lambda}(hh^{-1}xh)=h^{\blambda-\brho}f_{R,\lambda}(h^{-1}xh).
\ee
Now, using this formula and manifest expression (\ref{fR}), one finally gets
\footnote{In the following two formulas
we do not write down explicitly the subscript $I$ of the
Cartan torus coordinates $\bphi$ for more transparent formulas.}

\be
\pi_{\lambda}\left(e^{-\bmu_i\bphi T_{0,i}}\right)f_{R,\lambda}(x)=
e^{\left\{-\bmu_i\cdot\bphi\right\}\left\{(\blambda-\brho)\cdot
(\bfe_{i+1}-\bfe_i)\right\}}f_{R,\lambda}
(e^{\bmu_i\bphi T_{0,i}} xe^{-\bmu_i\bphi T_{0,i}})=\\=
e^{(\brho-\blambda)\bphi}f_{R,\lambda}(e^{\bmu_i\bphi T_{0,i}}
xe^{-\bmu_i\bphi T_{0,i}})=
e^{(\brho-\blambda)\bphi}e^{i\Tr x\mu
e^{\balpha\bphi}},
\ee
where the combination $\mu
e^{\balpha\bphi}$ means matrix with the matrix elements
$\delta_{i-1,j}\mu_ie^{\balpha_i\bphi}$. Thus, we finally get for LWF

\be\label{LWFIN}
\Psi(\phi)\sim e^{-\brho\bphi}\int _{X=B\backslash G} \prod_{i<j}dx_{ij}
h^{-(\blambda+\brho)}e^{(\brho-\blambda)\bphi}e^{i\Tr\mu x
e^{\balpha\bphi}}=\\=
e^{-\blambda\bphi}\int _{X=B\backslash G} \prod_{i<j}dx_{ij}
\prod_{i=1}^{N-1}\Delta_i^{-\2(\blambda\balpha_i+1)}(xx^t)
e^{i\mu_ix_{i,i+1}e^{\balpha_i\bphi}}.
\ee
In this formula we used that $(\blambda+\brho)(\bfe_{i+1}-\bfe_i)=
(\blambda+\brho)\balpha_i=\blambda\balpha_i+1$ and that
$\lambda$
is pure imaginary, and, therefore, $\overline
{f_{L,\lambda}}=f_{L,\bar\lambda}=f_{L,-\lambda}$.

\underline{Comment.}
Let us note that, in the $SL(3)$ case, we used the notations
$x_1$, $x_2$ and $x_{12}$ for $x_{12}$, $x_{23}$ and $x_{13}$ respectively.

\subsection{Gauss Whittaker LWF. Group derivation}
In order to get Gauss LWF we need to use instead of (\ref{rcglNIL})
condition (\ref{rcglNGL})
which gives $f_{L,\lambda}(x)$ (the right vector $f_R$ is certainly
the same). In the group form, this condition
is much analogous to (\ref{statecondRN}) and reads as
\be\label{statecondLNG}
\pi_{\lambda}(z^t) f_{L,\lambda}(g)=e^{iTr (\mu_L z)}f_{L,\lambda}(g).
\ee
To construct this function
we use the inner automorphism of the group $SL(N)$ which
maps strictly upper-triangular matrices to strictly lower-
triangular matrices. This automorphism can be manifestly described
as the matrix:
\beq
S=\left (
\begin{array}{ccccccc}
0 & 0      & 0      & 0     & 0  & 0      & 1 \\
0 & \ldots & 0      & 0     & 0  & 1      & 0 \\
0 & \ldots & 0      & 0     & 1  & 0      & 0 \\
0 & \ldots & 0      & 1    & 0  & \ldots & 0 \\
  & \ldots & 1      & 0     & 0  & \ldots &   \\
0 & 1      & 0      &\ldots & 0  & \ldots & 0 \\
1 & \ldots & 0      & 0     & 0  & \ldots & 0
\end{array}
\right )\ ,
\eeq
i.e.
\beq
S_{ij}=\delta _{i+j,N+1}
\eeq
and
\beq\label{tmat}
(S^{-1}zS)_{ij}=
z_{N+1-i,N+1-j}.
\eeq
Indeed, we need for the matrix $S$ to be an element of the $SO(N)$ group
(in general case we construct it as an element of the Weyl group, see
Part II).
Therefore, it should be normalized to the unit determinant. It can be done
by multiplying it by the factor -1 that does not contribute to formula
(\ref{tmat}). Now let us note that the automorphism $S$ reflects the element
$z_{i,i-1}$ into the element $z_{N-i,N+1-i}$ instead of $z_{i-1,i}$
(because of condition (\ref{rcglNGL}) we are only interested in those
elements of $z$ lying on the diagonal nearest to the main one). The situation
can be corrected by the suitable reflection of the matrix $\mu$. That is,
one needs to introduce the new matrix $\tilde\mu_L\equiv
\mu_{N-i}^L\delta_{i-1,j}$ so that condition (\ref{statecondLNG}) can be
rewritten as

\be\label{newcond}
\pi_{\lambda}(SzS^{-1})f_L(g)=f_L(gSzS^{-1})
e^{iTr (\tilde\mu_L z)}f_{L,\lambda}(gS^{-1}S),
\ee
i.e.
\be
f_L(gS^{-1}z)=
e^{iTr (\tilde\mu_L z)}f_{L,\lambda}(gS^{-1}).
\ee
From the consideration in the previous subsection we know the solution to this
equation, at least, in the upper-triangle matrices:
\be
\left.f_L(gS^{-1})\right|_{B_-=0}=f_L(n_+)=e^{iTr (\tilde\mu_L n_+)},
\ee
where $n_+$ is the upper-triangle part of $gS^{-1}$. Since we need to find out
the solution to (\ref{newcond}) for $g=x\in N_+$,
some recalculation is needed:
\be\label{1}
f_L(xS^{-1})=\chi_{\lambda}(xS^{-1})f_L(n_+)=\chi_{\lambda}(xS^{-1})
e^{iTr (\tilde\mu_L n_+)}.
\ee
Now we calculate this function in more explicit terms. First, analogously
to formula (\ref{fL}), diagonal part of $xS^{-1}$ is given by the ratios of
the corresponding minors $h_i={\Delta_i(xS^{-1})\over \Delta_{i-1}(xS^{-1})}$:

\be\label{2}
\chi_{\lambda}(xS^{-1})=\prod_i\Delta_i^{(\blambda\balpha_i-1)}(xS^{-1}).
\ee

To get the formula for the elements $(n_+)_{p-1,p}$ note that they depend only
on the $p\times p$ submatrix in the left upper corner. Let us
consider $(p-1,p)$ element of the matrix $b^{-1}\equiv n_+g^{-1}$, $g=xS^{-1}$.
It is equal to
zero by definition of $b$ and we have the identity:
\beq
(n_+)_{p-1,p}(g^{-1})_{p,p}+(n_+)_{p-1,p-1}(g^{-1})_{p-1,p}=0.
\eeq
Using the explicit expression for elements of the inverse matrix, we obtain
the desired expression:
\beq\label{3}
(n_+)_{p-1,p}=\frac {\Delta_{p-1,p}(xS^{-1})}{\Delta_{p-1}(xS^{-1})},
\eeq
where $\Delta_{p-1,p}(xS^{-1})$ is defined as the determinant  of the
$(p-1)\times (p-1) $ submatrix of $xS^{-1}$ with interchanged $p-1$ and
$p$ columns.

Thus, we have manifest expressions for $f_L$ and $f_R$. Collecting them
and making calculations completely analogous to those in formula
(\ref{LWFIN}) (see also (\ref{Cartan})), we get the final result for the Gauss
Whittaker LWF:

\be\label{res}
\Psi (\phi)=e^{-\blambda\bphi}
\int _{X=B\backslash G} \prod_{i<j}dx_{ij}\prod_{i=1}^{N-1}
\Delta_i^{-(\blambda\balpha_i+1)}(xS^{-1})\times\\\times
e^{ i\mu_{i}^Rx_{i,i+1}e^{\balpha_i\bphi} -i\mu_{N-i}^L
\frac{\Delta _{i,i+1}(xS^{-1})}{\Delta _{i}(xS^{-1})}}.
\ee

\subsection{Gauss Whittaker LWF. Algebraic derivation}
The same expression can be derived also in the "algebraic" way
(i.e. solving immediately conditions (\ref{rcglNGL}) and
(\ref{rcglNGR})) like it was
done for $SL(2)$ and $SL(3)$ cases. We demonstrate this by solving
the only non-trivial condition (\ref{rcglNGL}).

The algebra $SL(N)$ acts on functions as
\be\label{def}
a\cdot f(X) \equiv \left.\frac{d}{dt}f(Xe^a)\right|_{t=0}
\ee
for any generator $a\in SL(N)$.\\
Let us consider the following  functions:
\be\label{phi}
\phi_k(X) \equiv \frac{T_{+,k}\cdot\Delta_{N-k}(X)}{\Delta_{N-k}(X)}\;\;,
\;\;\;\;\;\;k= 1, \ldots , N\;\; .
\ee
Since the action of $T_{+,i}$ (see definition (\ref{def}) can be represented
as the differentiation
\be
T_{+,k} = \sum_{j=1}^{i}x_{jk}\frac{\partial}{\partial x_{j,k+1}}\;\; ,
\ee
it is clear that $T_{+,k}\cdot\Delta_k(X)$ is the minor which
can be obtained from
$\Delta_k(X)$ by the permutation of the $k$-th and $k-1$-th columns
of the matrix $X$.
It is easy to see that the left action of the whole Borel subgroup
$B_{-}\subset SL(N)$ leaves $\phi_k(X)$ invariant.
Now we need to calculate the right action of the subgroup $N_{-}$, i.e.
\be\label{action}
\phi_k(Xe^{tT_{-,j}}) \equiv
\left.\Bigl[(\Delta_{N-k}(Xe^{tT_{-,j}})\Bigr]^{-1}
\frac{d}{ds}\Delta_{N-k}\left(Xe^{tT_{-,j}}e^{sT_{+,k}}\right)
\right|_{s=0}\;\;.  \ee
The first factor in the r.h.s. of (\ref{action}) is right invariant
with respect to the action of $N_{-}$; all we need is to calculate the action
of $N_{-}$ in the second factor. Make use of the formula
\be
e^{tT_{-,j}}e^{sT_{+,k}} = \exp\Bigl(-\delta_{jk}T_{0,k}\log(1+st)\Bigr)
\times\\\times
\exp\Bigl(s(1+st\delta_{jk})T_{+,k}\Bigr)\cdot
\exp\Bigl(\frac{t}{1+st\delta_{jk}}T_{-,j}\Bigr)
\ee
Due to the right invariance of $\Delta_{N-k}$, the exponential containing
$T_{-,j}$ is dropped out. Thus,
\be
\Delta_{N-k}\left(Xe^{tT_{-,j}}e^{sT_{+,k}}\right) =
\Delta_{N-k}\Bigl(Xe^{-st\delta_{jk}H_k}e^{sT_{+,k}}\Bigr) +
O\bigl(s^2\bigr)\;.
\ee
Therefore, with the help of (\ref{weight}), equation (\ref{action})
reduces to
\be
\phi_k(Xe^{tT_{-,j}}) = \phi_k(X) + t \delta_{jk}
\ee
and, therefore, the function
\be
\Phi(X) \equiv \prod_{k=1}^{N-1}\exp\{i\mu_k\phi_k(X)\}
\ee
satisfies the conditions
\be
\Phi(B_{-}X) = \Phi(X) ,\
\Phi(Xe^{t_jT_{-,j}}) = e^{t_ji\mu_j}\Phi(X),\;\;\;\;\;\;j=1,\ldots , N-1\; .
\ee
Now, using this function, we can construct the function which belongs to
the representation induced by the character $\chi_{\blambda}$
and satisfies the condition (\ref{rcglNGL})
\be
T_{-,i}\cdot f_{\blambda}(X) = i\mu_i f_{\blambda}(X).
\ee
For doing this, we multiply $\Phi(X)$ by the proper degrees of $\Delta_k(X)$.
Then, using formulas (\ref{bor}) and (\ref{inv}),
one can finally find the solution
to (\ref{rcglNGL})
\be
f_{\blambda}(X) \equiv
\Phi(X)\prod_{k=1}^{N-1}\Bigl\{\Delta_k(X)\Bigr\}^{\blambda\balpha_k-1},
\ee
which coincides with the function $f_L$ given by (\ref{1}) with taking into
account (\ref{2}) and (\ref{3}).

\subsection{LWF by the Fourier transform}
As we could see in the previous examples, working in the Fourier
representation gives a very direct and transparent way. Looking at the results
obtained for the $SL(2)$ and $SL(3)$ cases, one could hope that there
existed the general formula for arbitrary $SL(N)$ expressed through
$\Gamma$-functions. However, it turns out to be not the case.
Indeed, let us consider the Fourier transformed LWF:
\be
f(\bphi)\equiv\int\prod d^{N-1}p e^{i\bp\bphi}\bar f(\bp),
\ee
where $d^{N-1}p$ is the volume element. The function $\bar f(\bp)$ satisfies
the following difference equation (we put again $\mu^L_i\mu^R_i=1$)

\be
-\2(\bp^2+\blambda^2)\bar f(\bp)=\sum_i\bar f(\bp+i\balpha_i).
\ee
To get some impression of what the result for LWF could be,
we consider the solution to the simpler
equation at $\blambda=0$ for $SL(4)$ case. Then, the equation is

\be
(p_1^2+p_2^2+p_3^2-p_1p_2-p_2p_3)\bar f(p_1,p_2,p_3)=\\=
\bar f(p_1+i,p_2,p_3)+\bar f(p_1,p_2+i,p_3)+\bar f(p_1,p_2,p_3+i),
\ee
where $p_i\equiv \bp\bmu_i$.
It is already not easy to solve this equation immediately. However, since we
know the integral representation for the LWF (see sect.3.7), we can read off
the result from the Fourier transform of formula (\ref{res}). It looks like
(compare with (\ref{sl3fg}))

\be
\int\ldots\int dx_1dx_2dx_3dx_{12}dx_{23}dx_{123}{\f x_{123}}\times \\\times
{\f x_{123}x_2-x_{12}x_{23}}{\f x_{12}x_3+x_1x_{23}-x_1x_2x_3-x_{123}}
\times \\\times e^{-x_1e^{\xi_1}-x_2e^{\xi_2}-x_3
e^{\xi_3}-{x_{12}\over x_{123}}-
{x_{123}-x_1x_{23}\over x_{123}x_2-x_{12}x_{23}}-
{x_{23}-x_2x_3\over x_{12}x_3+x_1x_{23}-x_1x_2x_3-x_{123}}}=\\=
\int\ldots\int {dx_1\over x_1} {dx_2\over x_2}{dx_3\over x_3}
{dt_{12}dt_{23}dt_{123}\over t_{123}(t_{123}-t_{12}t_{23})(t_{12}+t_{23}-1
-t_{123})}\times \\\times
e^{-x_1e^{\xi_1}-x_2e^{\xi_2}-x_3e^{\xi_3}-
{\f x_3}{t_{12}\over t_{123}}-
{\f x_2}{t_{123}-t_{23}\over t_{123}-t_{12}t_{23}}-
{\f x_1}{t_{23}-1\over t_{12}+t_{23}-1-t_{123}}}=\\=
\int\ldots\int {dx_1\over x_1} {dx_2\over x_2}{dx_3\over x_3}
{d\tilde x_1\over \tilde x_1}
{d\tilde x_2\over \tilde x_2}{d\tilde x_3\over \tilde x_3}
{dt_{12}dt_{23}dt_{123}\over t_{123}(t_{123}-t_{12}t_{23})(t_{12}+t_{23}-1
-t_{123})}\times \\\times dp_1dp_2dp_3
\left({x_3\tilde x_3 t_{123}e^{-\xi_3}\over t_{12}}\right)^{{p_3\over i}}
\left({x_2\tilde x_2 (t_{123}-t_{12}t_{23})e^{-\xi_2}\over
t_{123}-t_{23}}\right)^{{p_2\over i}}\times \\\times
\left({x_1\tilde x_1 (t_{12}+t_{23}-1-t_{123})e^{-\xi_1}\over
t_{23}-1}\right)^{{p_1\over i}}
e^{-x_1-x_2-x_3-\tilde x_1-\tilde x_2-
\tilde x_3}=\\=
\int\ldots\int \left\{\left[
dx_1dx_2dx_3d\tilde x_1d\tilde x_2d\tilde x_3x_1^{{p_1\over i}-1}
x_2^{{p_2\over i}-1}x_3^{{p_3\over i}-1}
\tilde x_1^{{p_1\over i}-1} \tilde x_2^{{p_2\over i}-1}
\tilde x_3^{{p_3\over i}-1}
\right.\right.\times \\\times\left.\left.
e^{-x_1-x_2-x_3-\tilde x_1-\tilde x_2-
\tilde x_3}\right]
dt_{123}dt_{12}dt_{23}\left[
t_{123}^{{p_3\over i}-1} t_{12}^{-{p_3\over i}}
(t_{123}-t_{12}t_{23})^{{p_2\over i}-1}
\right.\right.\times \\\times\left.\left.
(t_{123}-t_{23})^{-{p_2\over i}}
(t_{12}+t_{23}-t_{123}-1)^{{p_1\over i}-1}
(t_{23}-1)^{-{p_1\over i}}
\right]\right\}\times\\\times
e^{i\xi_1p_1+i\xi_2p_2+i\xi_3p_3}dp_1dp_2dp_3
\stackreb{{\hbox{Fourier}}\atop{\hbox{transf.}}}{\longrightarrow}
\Gamma^2({p_1\over i})\Gamma^2({p_2\over i})\Gamma^2({p_3\over i})
F({p_1\over i},{p_2\over i},{p_3\over i}),
\ee
where we introduced the new variables
$\displaystyle{t_{12}\equiv {x_{12}\over x_1x_2}}$, $\displaystyle{t_{23}\equiv
{x_{23}\over x_2x_3}}$, $\displaystyle{t_{123}\equiv {x_{123}\over x_1x_2x_3}}$
and the integral function
\be
F(\gamma_1,\gamma_2,\gamma_3)\equiv
\int\int\int dt_{123}dt_{12}dt_{23}
t_{123}^{\gamma_3-1} t_{12}^{-\gamma_3}\times\\\times
(t_{123}-t_{12}t_{23})^{\gamma_2-1}
(t_{123}-t_{23})^{-\gamma_2}
(t_{12}+t_{23}-t_{123}-1)^{\gamma_1-1}
(t_{23}-1)^{-\gamma_1}.
\ee
One can understand
that this is the general structure of the Fourier transformed LWF that, with
any simple root $i$, there can be associated $\Gamma^2({p_i\over i})$.
Unfortunately, there is no further factorization for non-simple roots.

The integral $F(\gamma_1,\gamma_2,\gamma_3)$
can be calculated, the result being the Meyer function \cite{BE}

\be
F(\gamma_1,\gamma_2,\gamma_3)=\int dxdydt x^{\gamma_2-1}(x-1)^{-\gamma_1-
\gamma_3+1}
\times\\\times (1-t)^{\gamma_3-1}t^{-\gamma_3}
(1-tx)^{\gamma_1-1}(1-y)^{p+1-1}y^{-\gamma_1}(1-yx)^{\gamma_3-1}=\\=
\Gamma(1-\gamma_3)\Gamma(\gamma_3)\Gamma(1-\gamma_1)\Gamma(\gamma_1)
\times\\\times\int dxx^{\gamma_2-1}
\phantom{F}_2F_1(\gamma_1,\gamma_3;1;x)\phantom{F}_2F_1(1-\gamma_3,1-
\gamma_1;1;x)=\\=
G^{33}_{44}\left(1\left|
\begin{array}{cccc}
1-\gamma_2, & \gamma_3, & \gamma_1, &1-\gamma_2\\
0, & \gamma_1-\gamma_2, & \gamma_3-\gamma_2, & 0
\end{array}\right.\right)\ ,
\ee
where $\displaystyle{x\equiv {t_{123}-t_{12}t_{23}\over t_{123}-t_{23}}}$,
$\displaystyle{t\equiv {t_{12}\over x}}$, $\displaystyle{y\equiv
{1-t_{23}\over x}}$, $\phantom{F}_2F_1(a,b;c;z)$
is the hypergeometric function and we
used formulas (\ref{hgir}), (\ref{hgpr}).  This function (at given values of
arguments) seems not to reduce to more elementary functions like
$\Gamma$-functions \cite{GR,BMP,BE}, although the Meyer function is defined to
be the Mellin transform of the ratio of products of $\Gamma$-functions (see
formula (\ref{irmf})), i.e.
\be
F(\gamma_1,\gamma_2,\gamma_3)\sim\int ds\Gamma(\gamma_1-\gamma_2+s)
\Gamma(\gamma_3-\gamma_2+s)
\Gamma(1-\gamma_3-s)\Gamma(1-\gamma_1-s).
\ee

The other possibility to use the Fourier transform is to do a canonical
transformation of variables and, then, proceed the calculations in the
momentum space (like it was done in sect.1.4.5). This results to the
formulas for the Iwasawa (\ref{LWFIN}) and Gauss (\ref{res}) integral
representations of LWF.

\subsection{Harish-Chandra functions}
Now let us discuss Harish-Chandra functions for the $SL(N)$ case. There are
two different points. The first point concerns the proper normalization, and
the second one is what are the asymptotics themselves. To fix the
normalization, one needs to remark that, because of the Weyl invariance of
the group theory construction (of the matrix elements), the LWF's are to be
Weyl invariant if properly normalized. In the Iwasawa case, this requires a
non-trivial normalizing factor, and we will discuss it in the next Part and
Appendix B. In the Gauss case, just integral (\ref{res}) turns out to
be Weyl invariant, by modulo some trivial power factors of $i$ and $\mu_j^R$.

Let us first look at the $SL(2)$ Gauss LWF (\ref{LWFG}). This integral is
invariant with respect to action of the Weyl group element $\lambda\to
-\lambda$ provided $\mu_R=i$. This condition removes the trivial non-invariant
factor $\left({i\over\mu_R}\right)^{-\lambda}$.
The invariance can be trivially demonstrated by the change of
variable in integral (\ref{LWFG}):
\be
x\longrightarrow {\mu_L\mu_Re^{-2\phi}\over x},
\ee
which leads to the same integral, with $\lambda$ replaced by $-\lambda$.

Of course, Weyl invariance of the integral can be immediately observed from
the Fourier transformed LWF (\ref{sl2F}). The invariance of the $SL(3)$
Gauss LWF can be also understood in the simplest way from the Fourier
transformed formula (\ref{fsl3}). However, in the previous section we
demonstrated that, in the case of higher rank groups, there would be no so
simple formula for the Fourier transformed LWF, and one has to use either
change of variables in the integrals, or more sophisticated methods in order to
prove Weyl invariance of the LWF. The most effective way to prove it
in general case is to construct operators that
intertwines between representations $\pi_{\blambda}$ and $\pi_{s\blambda}$,
where $s$ is an element of the Weyl group. Since this way is used in the
second Part (for the Iwasawa case), we say here some words on change of
variables in the Gauss LWF integrals showing up their Weyl invariance.

Let us consider the $SL(3)$ integral (\ref{psisl3}). Then, there are 6 elements
of the Weyl group. The distinguished one is the longest element $S$
($\lambda_1\to -\lambda_2$, $\lambda_2\to -\lambda_1$), and
its action to the integral can be described by the following change of
variables:
\be
\left\{
\begin{array}{l}
x_1\to {\f \mu^L_2\mu^R_2e^{\phi_2-2\phi_1}}{x_1\over x_{12}}\\
x_2\to {\f \mu^L_1\mu^R_1e^{\phi_1-2\phi_2}}{x_2\over x_1x_2-x_{12}}\\
x_{12}\to {\f \mu^L_1\mu^L_2\mu^R_1
\mu^R_2e^{-\phi_1-\phi_2}}{\f x_{12}}
\end{array}\right..
\ee
This can be easily generalized to the $SL(N)$ case (see comment at the end of
sect.3.6):
\be
x_{ij}\to
e^{-\balpha_{ij}\bphi}{\Delta_{i,j}(xS^{-1})\over\Delta_i(xS^{-1})},
\ee
where
$\balpha_{ij}$ denotes the positive root corresponding to $x_{ij}$ and we
omitted evident $\mu$-factors ($\mu^L_{N-i}\mu^R_{N-i}$ for $x_{i,i+1}$,
$\mu^L_{N-i}\mu^R_{N-i}\mu^L_{N-i-1}\mu^R_{N-i-1}$ for $x_{i,i+2}$ etc.).
Unfortunately, the other Weyl elements act less trivially (the element $S$ is
evidently distinguished in the Gauss LWF integral (\ref{res})!). Say, the Weyl
transformation $\lambda_1\to -\lambda_1$, $\lambda_2\to\lambda_1+\lambda_2$ in
the $SL(3)$ case is realized by the quite tedious change of variables
\be
\left\{
\begin{array}{l}
x_1\to -{\omega x_{12}+1\over\omega x_2x_{12}}(x_{12}-x_1-x_2)\\
x_2\to -{\omega x_{12}-1\over\omega x_{12}+1}x_2\\
x_{12}\to -{\f\omega}{x_{12}-x_1x_2\over x_{12}}
\end{array}\right.,
\ee
where we denoted for brevity $\omega\equiv\mu^L_1\mu^R_1e^{\phi_2-2\phi_1}$.

Now one can fix the normalization of the LWF by requiring that, first, it has
no poles, and, second, it is Weyl invariant. It still preserves the freedom of
multiplying by any Weyl-invariant polynomial. In order to fix normalization
entirely, one can ask "the minimal configuration", i.e. the absence
of any zeroes not fixed by the first two requirements\footnote{The same
normalization can be fixed in the very different ways
like it was done in footnote 6.} (this slightly
resembles the notorious CDD-ambiguities in $S$-matrix but, unlike them, the
ambiguities we discuss here do not influence the physical quantities like
$S$-matrix).

After fixing the normalization, one can calculate the asymptotics of the LWF,
i.e. Harish-Chandra functions. Since the number of different asymptotics
coincides with the number of elements of the Weyl group, the Harish-Chandra
functions are labeled by the Weyl group element, and connected by the action of
the Weyl group. In the Iwasawa case, the calculation of the asymptotics has
been done in paper \cite{GK} (see also the next Part). In the Gauss case, it is
done in the similar way, the result being (by modulo inessential $\mu$-factors)

\be
c_s(\blambda)=\prod_{\balpha\in\De^+}{\f \Gamma(1+s\blambda\cdot\balpha)},
\ee
where $s$ means an element of the Weyl group, and the product runs over all
the positive roots.
This coincides with the properly normalized Iwasawa formula -- see the next
Part
and Appendix B, how it had to be -- see s.3.5. This formula is also to be
compared with the Harish-Chandra functions obtained for the $SL(2)$ and $SL(3)$
groups (\ref{hchsl2}) and (\ref{hchsl3}) (see also comment on the affine case
in sect.2).

\newpage
\part{Construction for arbitrary group}
\setcounter{section}{0}
Now we are going to formulate the constructions of the previous Part
in the very general form, which though being much similar to that discussed
above (and sometimes following just the same line) still makes some sense,
since allows one to incorporate the Whittaker WF into more general
framework and to establish the connections with zonal spherical functions
and Calogero-Sutherland systems. On the other hand, this more general approach
also turns out to be useful in the affine case.

\section{General approach to arbitrary groups}
\setcounter{equation}{0}
Let $G$ be a semisimple  split Lie group. In particular, all complex groups are
split. $SL(n,{\bf R})$ is also the split group. Such groups have a Whittaker
model  of representations, which  will be described below. We will
work with the real forms. The reason why we use the split forms is that they
produce the most non-degenerate interactions of the Liouville type.
The real forms are  more convenient than the complex ones, since they allow
one to represent the generic relations and to draw an analogy with the
Calogero-Sutherland systems, which we plan to do.  The classical real split
groups are $A_{n-1}\rar SL(n,{\bf R}), ~B_n\rar SO(n+1,n),~C_n\rar Sp(n,{\bf
R}),~D_n\rar SO(n,n)$.

We start with some basic facts about the
representations of noncompact groups, which can be found
 in the textbooks \cite{Wa,He}. The most of relations will be valid for any
real forms, not only for the split ones, unless otherwise is specified.

\subsection{Cartan, Iwasawa  and Gauss decompositions}

Let ${\cal G}$ be a split real semisimple Lie algebra,
and $\si$ is the Cartan involution ($\si^2=id,~\si\neq id$). There is the {\em
Cartan
decomposition} of ${\cal G}$ in two eigensubspaces of $\si$ (the ${\bf Z}_2$
grading):
$${\cal G}={\cal K }+{\cal P},~\si{\cal K}={\cal K},~\si{\cal P}=-{\cal P}.$$
There exists such $\si$ that ${\cal K}$ is a maximal compact subalgebra in
${\cal G}$.
Let ${\cal A}$ be a  Cartan subalgebra
in ${\cal P}$. The split property of ${\cal G}$ means that $r=\dim {\cal A}=$
rank of ${\cal G}$ and ${\cal A}$ serves as a Cartan
subalgebra for ${\cal G}$. There is one real split form for any simple complex
Lie algebra ${\cal G}_{\bf C}$. A group is split if its algebra
is split.

Let $\{\al\}=\De$ be the root system in the
dual space ${\cal A}^*$. It means that
$${\cal G}={\cal G}_0~\oplus_{\al}{\cal G}_{\al},$$
$${\cal G}_{\al}=\{x\in{\cal G}~|~{\ad}_hx=\al(h)x, ~h\in {\cal A}\}$$
and ${\cal G}_0={\cal A}+{\cal M},~{\cal G}_0\cap{\cal K}={\cal M}$.
$\dim {\cal G}_{\al}=m_{\al}$ is called the multiplicity of the root space
${\cal G}_{\al}$. For the split forms $m_{\al}=1$ and ${\cal M}=\emptyset$.

 The involution
acts on the root subspaces as
$$\si{\cal G}_{\al}\ms -{\cal G}_{-\al}.$$
Therefore,
$$
{\cal K}=\oplus_{\al}({\cal G}_{\al}-{\cal G}_{-\al}),
$$
and
$${\cal P}={\cal A}+\oplus_{\al}({\cal G}_{\al}+{\cal G}_{-\al}).$$

The Killing form $<$ , $>$ on ${\cal G}$ is non-degenerate
and positive definite on $P$.
It takes the canonical Euclidean form on ${\cal A}$ and
$$<{\cal G}_{\al},{\cal G}_{\bet}>=\de_{\al,-\bet}.$$

There are two important facts about the Cartan decomposition:

\bigskip

\noindent
i) Generic $x\in{\cal P}$ can be "diagonalized" by the adjoint action of $K$:
 $x=\Ad_k\phi,~\phi\in  {\cal A},~k\in K$.
The element $\phi$ is defined up to the action
of the Weyl group $W=M'/M$ ($M'(M)$ is the normalizer (the centralizer)
of $A$ in $K$). Therefore,
$a$ can be taken lying in the Weyl chamber $\La=A/W$.\\
ii) Cartan decomposition can be lifted to the polar decomposition
$$
G=PK,~ P=\exp{\cal P}.
$$

\bigskip

\noindent
Here $P=G/K$ is a symmetric space of noncompact type and $\exp$ is one-to-one
mapping $\exp:{\cal P}\rar P$. Therefore, for generic $g$
$$
g=k_1ak_2,~k_j\in K,~\log a\in\La\subset{\cal A}\subset {\cal P}.
$$

We fix some  ordering
in  the dual space ${\cal A}^*$. It simply means choosing the hyperplane in
${\cal A}^*$ which does not contain any root. It divides ${\cal A}^*$
to the positive and negative parts.
 Let $\De^+$ be a subsystem
of the positive roots with respect to this ordering $(\De=\De^+\cup \De^-)$ .
It allows one to specify the positive
Weyl chamber $\La^+=\{a\in{\cal A}~|~ \al(a)>0~\forall \al\in\De^+\}$.
The simple roots $\al\in\Pi\subset\De^+$ generate a basis in $\De^+$;
arbitrary $\al\in\De^+$ is a sum of the simple roots with non-negative integer
coefficients.

 Let ${\cal N}$ be  a nilpotent subalgebra generated by the positive root
subspaces:
$${\cal N}=\oplus_{\al\in\De^+}{\cal G}_{\al}.$$
The algebra ${\cal G}$ can be represented as the direct sum of its
subalgebras
 $${\cal G}={\cal K }+{\cal A}+{\cal N}.$$
It is called the {\em Iwasawa
decomposition}.
This decomposition can be lifted to the corresponding group element
decomposition
$$\label{I1}
G=KAN.\hspace{3cm}({\rm I}1)
$$
Introduce the vector
$\rho\in{\cal A}^*~\rho=\frac{1}{2}\sum_{\al\in\De^+}m_{\al}\al.$
Here $m_{\al}=\dim {\cal G}_{\al}$ is the multiplicity of $\al$. For the real
split
group $m_{\al}=1$ and we come to the conventional formula for $\rho$.
Since $\Ad_a{\cal
G}_{\al}=e^{\al(\log a)}{\cal G}_{\al}$,
$$
\tr\Ad_a|_{\cal N}=2\rho(\log a).
$$

For technical reasons we will use also another Iwasawa decomposition as well.
Let $\bar{N}=\si (N)$.
It means that its Lie algebra is generated by the negative root subspaces.
Then we can represent the group as
$$\label{I2}
G=\bar{N}AK.\hspace{3cm}({\rm I}2)
$$
We call the element $h_I(g)$ coming from these  decomposition $g=vh(g)k$
the horospheric projection. We can define as well the other coordinates of
$g$: $\bar{n}(g)$ and $k(g)$. Note that $k(g)$ is defined up to the left
multiplication on $M$ (the centralizer of $A$ in $K$).  In general, we have
$$
k(gu)=k(g)u,~u\in K, ~~k(gm)=k(g), ~m\in M,
$$
\beql{g1}
h_I(vg)=h_I(g),~v\in {\bar N},
\eq
$$
h_I(av)=a,~\bar{n}(av)=ava^{-1},~a\in A,v\in {\bar N}.
$$
Generic group elements have also {\em the Gauss representations}.
The Gauss decomposition is
$$G=\bar{N}AN,$$
$$g=vhn=bn,~n\in N,~h\in A,~v\in {\bar N},~b\in B,$$
and $B$ is a Borel subgroup. The Gauss coordinates of $g$ are uniquely defined.
We will denote them as $h_G(g), n(g), \bar{n}(g)$. It  is clear that
\beql{g2}
h_G(rv)=h_G(yr)=r,~~n(yr)=\Ad^{-1}_r y,~~y\in N,~r\in A,~v\in {\bar N}.
\eq
The generalization of the Gauss decomposition is the Bruhat decomposition,
which covers the whole $G$. It allows one to define the Schubert
 cell decomposition of the flag variety $B\sm G$.
 The set $Y\sim  N$ is a  cell of maximal dimension in $B\sm G$. On the
other hand, due to the Iwasawa decomposition (\ref{I2}) the flag variety has
the representation $$B\sm G\sim M\sm K,$$
\beql{g7}
n\rar Mk(n),
\eq
which is the
diffeomorphism of $\bar{N}$ onto open the
dense subset of $ M\sm K$.  For
$SL(2,{\bf C})$,
it is the usual stereographic projection ${\bf C}\rar {\bf CP}^1$.  The
group $G$ acts on the flag variety as a group of diffeomorphisms
\beql{g3}
xg\mapsto k(xg), ~x=Mk,~ xg=\bar {N}AMk(xg),
\eq
$$yg\mapsto n(yg),~y\in N,~yg=\bar{N}An(yg).$$

Consider now integral formulas on $G$ related to the Iwasawa
decomposition (\ref{I2}) and to the
Gauss decomposition. Since $G$ is semisimple,
 there exists a measure $dg$,
which is both left and right invariant.
To derive $dg$, note that it can be reconstructed from
the invariant metric, which, in its turn, is defined by the Killing form
$$\ga=<g^{-1}\de g ,g^{-1}\de g >.$$
The invariant measure in local coordinates $x_1,\ldots,x_d,~d=\dim G$
takes the form
$$\omega=(\det\ga)^{1/2}dx_1\ldots dx_d.$$
If $g=vak$ (I2), then
$$<g^{-1}\de g ,g^{-1}\de g >=$$
$$<a^{-1}\de a,a^{-1}\de a>+<\de kk^{-1},\de kk^{-1}>
+<\Ad^{-1}_a v^{-1}\de v,k^{-1}\de k>.$$
 The non-zero components of the metric in the Iwasawa coordinates
take the form
$$
\ga_{i,j}=\de_{i,j}~{\rm for~the~canonical~ basis~}H_j~{\rm in}~{\cal A},
$$
$$
\ga_{i,j}=-{\rm Killing~ metric~on~}K,
$$
$$
\ga_{i,j}=2\exp\al(\log a),~i=\al,j=-\al.
$$
The determinant of the metric is equal
$$
\det\ga_{a,b}=\exp 4\rho(\log a).
$$
The measure $dg$ can be normalized in such a way
that for $g=vak$ and left-invariant measures $dv,da,dk$
$$
\int_Gdgf(g)=\int_{\bar{N}\times A\times K} dvdadkf(vak)e^{2\rho(\log a)}.
$$

This relation allows one to define an invariant measure $dx$ on $M\sm K$
such that $\int_{M\sm K}dx=1$ and under the right group action it is
transformed as
\beql{g5}
\int_{M\sm K}f(xg)h_I^{-2\rho}(gx)dx=\int_{M\sm K}f(x)dx,
\eq
where we use the notation $h_I^{-2\rho}(xg)=\exp(-2\rho(\log h_I(xg))).$

The metric in the Gauss coordinates $g=van,~v\in\bar{N}$ takes the form
$$
<g^{-1}\de g ,g^{-1}\de g >=<a^{-1}\de a,a^{-1}\de a>+
+2<\Ad_a\de nn^{-1}v^{-1}\de v>.
$$
It amounts
$$
\ga_{i,j}=\de_{i,j}~{\rm in~ the~ canonical~ basis ~}H_j~{\rm in~}{\cal A},
$$
\beql{g6}
\ga_{i,j}=\exp\al(\log a),~i=\al,j=-\al.
\eq
The determinant of the metric is equal to the determinant
 in the Iwasawa coordinates
$$
\det\ga_{a,b}=\exp 4\rho(\log a).
$$
It leads to the Gauss integral formula
$$
\int_Gdgf(g)=\int_{\bar{N}\times A\times N} dvdadnf(van)e^{2\rho(\log a)}.
$$
 The function $h_I^{-2\rho}(n)$ is integrable on $N$.
If the Haar measure on $M\sm K$
is normalized as above and
$$
\int_{N}dnh_I^{-2\rho}(n)=1,
$$
then
\beql{g8}
\int_{M\sm K}f(Mk)dk_M=\int_{N}f(Mk(n))h_I^{-2\rho}(n)dn.
\eq

\subsection{Principle series of irreducible representations}

Consider the Borel subalgebra $B=\bar{N}AM$
in $G$ and its character
$$
\chi_{\nu}(vam)=\exp(i\nu-\rho)(\log a),~m\in M,~v\in \bar{N},~a\in A.
\footnote{In principle, this character can be non-trivial on the
centralizer $M$ as well.
For our construction, it is sufficient to restrict it onto the Cartan
subgroup $A$ only.}
$$
Consider the space of smooth functions  on $G$
which satisfies the following relation
$$
f(bg)=\chi_{\nu}(b)f(g),~b\in B.
$$
The principle series of representations $\pi_{\nu}$ is defined as
$$
\pi_{\nu}(g)f(x)=f(xg).
$$
According to the  Borel-Weyl theorem, $\pi_{\nu}$
is realized in the space $\G_{\nu}$ of holomorphic sections
of the line bundle
${\cal L}_{\nu}$ over the flag variety $ B\sm G$.
There exists a Hermitian scalar product in $\G_{\nu}$ for $\nu\in {\bf R}$
$$
<\Psi_L|\Psi_R>=\int_{ B\sm G}\Psi_L\cdot\bar{\Psi}_R.
$$
Consider the realization of $ B\sm G$ as $ M\sm K$. Then, taking
into account the action of $G$ on $ M\sm K$ (\ref{g3}) and to the Iwasawa
decomposition (I2),
\beql{g9}
\pi_{\nu}(g)|\Psi>(x)=h_I^{i\nu-\rho}(xg)|\Psi>(k(xg)).
\eq
Due to (\ref{g5}) the representation
is unitary.
We call this realization the compact one.

Similarly, for the realization on $N$ the same representation
takes the form
\beql{g10}
\pi_{\nu}(g)|\Psi>(x)=h_G^{i\nu-\rho}(xg)|\Psi>(n(xg)).
\eq
This realization will be referred to as the noncompact one.

 The action  $\pi_{\nu}(g)$ of $G$ on the
left  vectors in
the space $\G_{\nu}$ in the both realizations is given by the
inverse operator $<\Psi_L|\pi_{\nu}(g^{-1})$, as it should be.

\subsection{Casimir and Laplace-Beltrami operators}

Consider the second order Casimir operator $C_2$ in the universal
enveloping algebra $U({\cal G})$. In the basis in ${\cal G}$
we have introduced
$\{H_j,{\cal G}_{\pm\al},~j=1\ldots,r,~\al\in\De^+\}$ it
takes the form
$$
C_2=\sum_{j=1}^rH_j^2+\sum_{\al\in\De^+}
{\cal G}_{\al}{\cal G}_{-\al}+{\cal G}_{-\al}{\cal G}_{\al}.
$$
Taking into account the commutation relations
$$
[{\cal G}_{\al},{\cal G}_{-\al}]=\al_j(H_j)H_j,
$$
we rewrite $C_2$ as
\beql{g10a}
C_2=\sum_{j=1}^r\left(H_j^2+2\rho(H_j)H_j\right)+
2{\cal G}_{-\al}{\cal G}_{\al}.
\eq
Being restricted to the irreducible representations, the Casimir operators act
 as scalars.
To calculate the value of $C_2$ acting  on $\G_{\nu}$ we remind that
the representation $\pi_{\nu}$ has the highest weight vector
$\xi_{\nu}$. For example, in
the compact realization ({\ref{g9}) it is a constant function
$$
\pi_{\nu}({\cal G}_{\al})\xi_{\nu}=0,~\al\in\De^+
$$
and
$$
\pi_{\nu}(H_j)\xi_{\nu}=(i\nu-\rho)(H_j)\xi_{\nu}.
$$
Acting by $\pi_{\nu}(C_2)$ on $\xi_{\nu}$, we find
$$
\pi_{\nu}(C_2)=-<\nu,\nu>-<\rho,\rho>.
$$

We need the explicit form of $C_2$ in the regular representation
$$f(x)\mapsto f(xg)$$
 in the
Iwasawa (I1) and Gauss coordinates. The derivation is based on the observation
that the second order Casimir  coincides  with
the Laplace-Beltrami operator $B$, constructed by means of the Killing metric
$\ga_{i,j},~i,j=1,\ldots,\dim G$ on $G$
\beql{g11}
B=\frac{1}{(\det\ga)^{1/2}}\p_i\ga^{i,j}(\det\ga)^{1/2}\p_j,
\eq
where $\ga_{i,j}\ga^{j,k}=\de_i^k$.  Note that this metric
is both left and right invariant
providing the invariance of the Laplace-Beltrami operator.
Let us calculate   the metric in  the Iwasawa coordinates (\ref{I2}).
If $g=kan$, then
\be
<g^{-1}\de g ,g^{-1}\de g >=<a^{-1}\de a,a^{-1}\de a>+
\\+<k^{-1}\de k,k^{-1}\de k>
+<\Ad_a\de nn^{-1},k^{-1}\de k>.
\ee
The non-zero components of the metric in the Iwasawa coordinates
takes the form
$$
\ga_{i,j}=\de_{i,j}~{\rm for~the~canonical~ basis~}H_j~{\rm in}~{\cal A},
$$
\beql{g4}
\ga_{i,j}=-{\rm Killing~ metric~on~}K,
\eq
$$
\ga_{i,j}=-2\exp\al(\log a),~i=\al,j=-\al.
$$
The determinant of the metric is equal
$$
\det\ga_{a,b}=\exp 4\rho(\log a).
$$

Let
$$a^{-1}\de a=\sum _{j=1}^{r}\phi_jH_j,~
\de nn^{-1}=\sum_{\al\in\De^+}n_{\al}{\cal G}_{\al},$$
$$k^{-1}\de k=\sum_{\al\in\De^+}k_{\al}({\cal G}_{\al}-{\cal G}_{-\al}).$$
Then, substituting (\ref{g4}) into (\ref{g11}), we find $B$ in
these local  coordinates
\beql{g12}
B_I=e^{-2\rho(\phi)}\sum_{j=1}^{r}\p_{\phi_j}e^{2\rho(\phi)}\p_{\phi_j}-
2\sum_{\al\in\De^+}\left[e^{-\al(\phi)}\p_{n_{\al}}\p_{k_{\al}}-
e^{-2\al(\phi)}\p^2_{n_{\al}}\right].
\eq
In the same way, the Gauss coordinates lead to the operator
\beql{g13}
B_G=e^{-2\rho(\phi)}\sum_{j=1}^{r}\p_{\phi_j}e^{2\rho(\phi)}\p_{\phi_j}+
\sum_{\al\in\De^+}e^{-\al(\phi)}\p_{n_{\al}}\p_{v_{\al}},
\eq
where $v^{-1}\de v=\sum_{\al\in\De^+}v_{\al}{\cal G}_{-\al}$.

For the completeness, we write down the Laplace-Beltrami operator in
the coordinates corresponding to the Cartan decomposition. Since in the
decomposition $g=k_1\exp \phi k_2$
the measure depends only on
the radial part $\phi$
$\det \ga=\prod_{\al\in\De^+}\sinh^{m_{\al}}\al(\phi)=\de (\phi)$,
\beql{g14}
B_{Cartan}=
\frac{1}{\de^{1/2} (\phi)}\sum_{j=1}^r\p_{\phi_j}\de^{1/2} (\phi)\p_{\phi_j}
+K{\rm~ dependent~"angular"~operator.}
\eq

Consider the higher Casimir operators.
There is a generalization of the representation of $C_2$
in the Iwasawa
coordinates \cite{HC}.
Let $u$ be an element from the universal enveloping algebra $U({\cal G})$.
Then there exists a unique element $p(u)\in U({\cal A})$ such that
$$u-p(u)\in {\cal K}U({\cal G})+U({\cal G}){\cal N}.$$
Define the map $\ga:~U({\cal G})\rar  U({\cal A}),~~\ga(u)=p(u)-H_{\rho}(u)$,
where $H_{\rho}$ is defined by the Killing metric on ${\cal A}$ $<H_{\rho},H>=
\rho(H)$.
The map $\ga$ induces  a homomorphism of the center
$Z({\cal G})$  of $U({\cal G})$
 (the algebra of the Casimirs) on the algebra of $W$-invariant polynomials
$I^W$ on ${\cal A}$
\beql{g15}
\ga:~~Z({\cal G})\rar I^W({\cal A}).
\eq
Applying this homomorphism to $C_2$ (\ref{g10a}), one
obtains $\ga(C_2)=-<H,H>-<\rho,\rho>$.

\subsection{Spherical vectors and zonal spherical functions}

It is important to find out such spaces of the irreducible representations
that there exist vectors invariant
with respect to the compact subgroup $K$ and covariant
with respect to the nilpotent subgroup
$\bar{N}$.

The representations containing vectors $<\Psi^K|,~|\Psi^K>$ invariant with
respect to $K$
are called the {\em spherical} ones.
The principle series of representations described above
are spherical.
The matrix elements $\Phi_{\nu}(g)=<\Psi^K|\pi_{\nu}(g)|\Psi^K>$
are called {\em zonal
spherical functions} (ZSF). It follows from their definition that they
depend only on the radial part $r(g)$ in the Cartan decomposition
$g=k_1r(g)k_2$.
Moreover, $\phi=\log r(g)$ can be chosen to lie in the positive Weyl chamber
$\phi\in\La^+$ and $\Phi_{\nu}(\phi)$ is $W$-invariant
$\Phi_{\nu}(s\phi)=\Phi_{\nu}(\phi)$.

Since this matrix element is defined in the irreducible representation,
$\Phi_{\nu}(\phi)$ is the common eigenfunction of all Casimir operators
$$C\Phi_{\nu}(g)=\ga(\nu)\Phi_{\nu}(g),$$
where $\ga(\nu)$ is determined by (\ref{g15}).
In particular, for $C_2$, writing it as the Laplace-Beltrami operator in the
Cartan coordinates (\ref{g14}) and using the $K$-bi-invariance
of the matrix element,
we come to the equation
$$\sum_{j=1}^r\left[\p^2_j+\sum_{\al\in\De^+}m_{\al}\coth\al(\phi_j)\p_j\right]
\Phi_{\nu}(\phi)
=-(<\nu,\nu>+<\rho,\rho>)\Phi_{\nu}(\phi).$$
After the gauge transform
$$\Phi_{\nu}(\phi)\mapsto
\psi_{\nu}(\phi)=\frac{\Phi_{\nu}(\phi)}{\de^{1/2} (\phi)},$$
we come to the eigenvalue problem for the Calogero-Sutherland system
\cite{OP1}
$$\left[-\sum_{j=1}^r\p^2_j+\sum_{\al\in\De^+}
\frac{m_{\al}(m_{\al}-1)}{2<\al,\al>}
\frac{1}{\sinh^2\al(\phi)}\right]\psi_{\nu}(\phi)=E_{\nu}\psi_{\nu}(\phi).
$$
Independently, ZSF $\Phi_{\nu}(\phi)$ can be uniquely fixed by the three
properties:

\bigskip

\noindent
i) It is a common eigenfunction of the Casimir operators with the
eigenvalue $\ga(\nu)$ (\ref{g15});\\ ii) It is a $K$-bi-invariant function on
$G$, and, therefore,  it lives on the double coset space $K\setminus G/K\sim
\La$;\\ iii) $\Phi_{\nu}(\phi)|_{\phi=0}=1.$

\bigskip

\noindent
It can be proved that the ZSF as the
functions of $\nu$ are $W$-invariant
\beql{g15a}
\Phi_{s\nu}(\phi)=\Phi_{\nu}(\phi),~s\in W.
\eq
Let us write down explicitly the integral representation for ZSF
(the Harish-Chandra
formulas \cite{HC}), and, thereby, for the wave functions of the
Calogero-Sutherland system. The compact case is the simplest one,
since the invariant state is just a constant function:
$$
\Phi_{\nu}(g)=\int_{M\sm K}dxh_I^{i\nu-\rho}(xg).
$$
Note that, instead of $g$, we can use here the element $r(g)=\exp(\phi)\in A$
from the Cartan decomposition $g=k_1rk_2$.

The noncompact case can be treated similarly, or, equivalently, one
can use the generalized
stereographic map $Mk\rar v\in N$ (\ref{g7}),(\ref{g8}).
It can be proved that the $K$-invariant
states in the noncompact realization are
$$<\Psi^K_L|(y)=h^{i\nu-\rho}_I(y),~y\in N.$$
It follows from the equality
$$h_I(n(yu))=h_I(y)h_G^{-1}(yu),~u\in K,~y\in N,$$
which can be derived by comparing the Gauss and the
Iwasawa decompositions (I2) for $y$, using (\ref{g1}).
Taking into account that
 $$
h_G(yr)=r,~n(yr)=r^{-1}yr,~r\in A
$$
and using (\ref{g10}), we come to the expression
\beql{g16a}
\Phi_{\nu}(r)=r^{i\nu-\rho}\int_Ndyh_I^{i\nu-\rho}(r^{-1}yr)
h_I^{-i\nu-\rho}(y).
\eq
Let
\beql{g16}
D=\{\nu\in{\cal A}^*|\Im m\nu_{\al}<0,~\forall \al\in\De^+\}.
\eq
Consider  $r=r(t)=\exp tH,~ H\in\La^+,~t\rar +\infty$.
Since $\Ad^{-1}_{r(t)}y\rar id$ and integral converges for $\nu\in D$ in this
limit, $$\Phi_{\nu}(r(t))\sim r^{i\nu-\rho}c(\nu).$$
Here $c(\nu)$ is the Harish-Chandra function
\beql{g17}
c(\nu)=\int_{\bar{N}}dyh_I^{-i\nu-\rho}(y).
\eq
This integral was calculated explicitly by the
factorization procedure (Gindikin-Karpelevich formula \cite{GK}).
$$
c(\nu)=\prod_{\al\in\De^+}c_{\al}(\nu),~c_{\al}(\nu)=I_{\al}(i\nu)/I_{\al}
(\rho),
$$
where for the split groups
\beql{g18}
I_{\al}(\nu)=B\left(\frac{1}{2},\nu_{\al}\right),~~
\nu_{\al}=\frac{{\al}(\nu)}{<\al,\al>}.
\eq
The Harish-Chandra function defines the scattering in the Calogero-Sutherland
model, since for $\log r(t)\in \La^+,~\nu\in{\cal A}^*,t\rar +\infty$
\beql{sc}
\Phi_{\nu}(r(t))\sim \sum_{s\in W}c(s\nu)r^{si\nu-\rho}.
\eq

 The scattering process is reduced to the two particle collisions due to
the Gindikin-Karpelevich formula.

\section{Whittaker model.}
\setcounter{equation}{0}
Here we present
the general theory of the Whittaker models. Their concrete realizations for
particular groups were given in Part I.

Consider one-dimensional representation $\psi_{\mu}$ of the group $\bar{N}$:
$$\psi_{\mu}(v_1v_2)=\psi_{\mu}(v_1)\psi_{\mu}(v_2),~\mu\in{\cal A}^*.$$
It can be constructed as follows. Represent $v$ as an element
of a matrix group $\bar{N}$ as $v=v'+\ti{v},$
where $\ti{v}\in [\bar{N},\bar{N}]$. The element $v'$ can be decomposed as
$v'=\sum_{\al\in\Pi}v_{\al},~v_{\al}\in{\cal G}_{\al}$. Since $v_1v_2=
v_1'+v_2'+\tilde{v}$,
\beql{woa}
\psi_{\mu}(v)=\exp i\mu(v')=\exp i\sum_{\al\in\Pi}\mu_{\al}v_{\al}.
\eq
The group $N$ has similar representations.

Consider the space $C^{\infty}_{\nu,\mu}$ of smooth functions on $G$,
which can be characterized  by the following properties:

\bigskip

\noindent
i) $V_{\nu}(g;\mu) $ are common eigenfunctions of the Casimir operators
with the eigenvalues $\ga(\nu)$;\\
 ii$^I)~V_{\nu}(kgn;\mu)=\psi_{\mu}(n)V_{\nu}(g;\mu),~n\in N,k\in K$
(the covariance condition).

\bigskip

\noindent
This repeats the corresponding conditions for the ZSF.
Let $V_{\nu}(g;\mu)\in C^{\infty}_{\nu,\mu}$. Then, taking $g=k(\exp \phi) n$,
we find from (\ref{g12}) that it satisfies the equation
\beql{w0}
\left[e^{-2\rho(\phi)}\sum_{j=1}^{r}\p_{\phi_j}e^{2\rho(\phi)}\p_{\phi_j}-
2\sum_{\al\in\Pi}\mu_{\al}^2e^{-2\al(\phi)}\right]V_{\nu}(\phi;\mu)=\\=
-(\nu^2+\rho^2)V_{\nu}(\phi;\mu).
\ee

Making substitution $V_{\nu}(\phi;\mu)=
\exp (-\rho(\phi))\vf_{\nu}(\phi;\mu)$,
one obtains
\beql{w1}
\left[
\sum_{j=1}^{r}\p^2_{\phi_j}-2\sum_{\al\in\Pi}\mu_{\al}^2e^{-2\al(\phi)}\right]
\vf_{\nu}(\phi;\mu)=-\nu^2\vf_{\nu}(\phi;\mu).
\eq
It is just the eigenvalue problem for the Schr\"odinger operator for the
open Toda lattice.

Let $L$ be a root sublattice
$$L=\sum_{\al\in\Pi}n_{\al}\al,~n_{\al}~
{\rm are~even~ integer~and~non-negative.}$$
Define  rational functions $a_{\ga}(\nu)$ on ${\cal A}^*$
depending on the vertices $\ga$
of $L$ by the recurrence relation
$$
(-\nu^2+2<\nu,\ga>)a_{\ga}(\nu)-
2\sum_{\al\in\Pi}\mu_{\al}^2a_{\ga-2\al}(\nu)=0,~a_{0}(\nu)=1.
$$
Then, it can be demonstrated by direct calculations  that
 \beql{w2}
\vf_{\nu}(\phi;\mu)=e^{i\nu(\phi)}\sum_{\ga\in L}a_{\ga}(\nu)e^{-\ga(\phi)}.
\eq
Let $\tilde {\cal A}^*$
be the complement in ${\cal A}^*$ to the union of hyperplanes\\
$\si_{\nu}=\{-\nu^2+2<\nu,\ga>=0\}$. Then, the series converges absolutely
and uniformly for $\phi\in{\cal A},~\nu\in \tilde{\cal A}^*$ \cite{Ha}.
Moreover, it was proved that
$$V(r;s\nu,\mu),~s=id,\ldots,s_0,~s\in W$$
generate a basis in $C^{\infty}_{\nu,\mu}(A),~\dim C^{\infty}_{\nu,\mu}(A)=
\sharp W$ . It is easy to see that
 they are unbounded on $A$.

Of our main interest,
however, are bounded functions in this space. In fact, there
exists only one such a
function in $C_{\nu,\mu}(A)$. It is precisely the function
we call {\em the Whittaker function} (WF).
Following the same ideology as in the ZSF case, we construct WF's
as matrix elements in the irreducible space $\G_{\nu}$.
We also add to i), ii) the normalization condition which is not so
evident as iii) for the ZSF. We work in the noncompact realization.
It follows from the covariant condition that
we should define the state $w(y;\mu)=|\Psi^{\mu}_R>$ such that
$$
\pi_{\nu}(n)|\Psi^{\mu}_R>(y)=\psi_{\mu}(n')|\Psi^{\mu}_R>(y),~
(n=n'+\ti{n},~\ti{n}\in [N,N]).
$$
It is clear that $w(y;\mu)=\psi_{\mu}(y)=\exp i\mu(y')$. This state is called
{\em the Whittaker vector}. There is only one Whittaker vector in $\G_{\nu}$ up
to the constant multiplier. We have already defined
the  $K$-invariant states
in the noncompact realization: \\ $<\Psi^K_L|= h_I^{i\nu-\rho}(y)$. The matrix
element $<\Psi^K_L|\pi_{\nu}(g)|\Psi^{\mu}_R>$ satisfies
both conditions i) and ii). Thus, using (\ref{g1}), we find
\beql{w3}
W_{\nu}(r;\mu)=r^{-i\nu+\rho}\int_Ndyh_I^{i\nu-\rho}(ryr^{-1})
\exp i\mu(y').
\eq
The integral converges for $\nu\in{\cal A}^*_{\bf C},~\Re e\nu\in\La$
\cite{Ha}. Therefore,
we have constructed the WF -- the desirable bounded eigenfunction. This
expression is close to the similar integral for the ZSF. It can be rewritten in
another form, if one considers the group action on the right state. Using
(\ref{g2}), we obtain \beql{w4}
W_{\nu}(r;\mu)=r^{-i\nu-\rho}\int_Ndyh_I^{i\nu-\rho}(y)
\exp i\mu(r^{-1}yr)'.
\eq
Like the ZSF case assume that
$r=r(t)=\exp tH,~ H\in\La^+,~t\rar +\infty$ $\Ad_{r^{-1}}y=id$. Then, we
obtain from (\ref{g17}) and (\ref{w4}) for
$-\nu\in D$ (\ref{g16})
\beql{w5}
\li r^{i\nu+\rho}W_{\nu}(r(t);\mu)= c(-\nu).
\eq
It will be instructive to compare these integral representations with
the similar formulas for $G=SL(2,{\bf R})$ in Part I. In this case, we have
$y=y'\in {\bf R},~r\in {\bf R}^+$
$$
h_I(y)=\di\left((1+y^2)^{\2},(1+y^2)^{-\2}\right),~~(r^{-1}yr)'=yr^{-2},~
$$
$$
i\nu-\rho=\di\left(i\nu-\2,-i\nu+\2\right).
 $$
Putting $r=e^{\phi}$ we come to the integrals (compare with (I.\ref{LWFI}))
$$
W_{\nu}(\phi;\mu)=e^{(-2i\nu+1)\phi}\int_{-\infty}^{\infty}\frac
{\exp(-i\mu y)}
{(1+y^2e^{4\phi})^{-i\nu+\2}}dy,
$$
$$
W_{\nu}(\phi;\mu)=e^{(-2i\nu-1)\phi}\int_{-\infty}^{\infty}\frac
{\exp(-i\mu ye^{-2\phi})}
{(1+y^2)^{-i\nu+\2}}dy.
$$
The direct calculation gives
$$
W_{\nu}(\phi;\mu)=\left(\frac
{2}{\mu}\right)^{i\nu}
\frac{1}{\G(\2-i\nu)}e^{-\phi}K_{-i\nu}(\mu e^{-2\phi})
$$
as it should be.
Now we present the expansion of the WF in the basis
$V(r;s\nu,\mu),~s\in W$ \cite{Ha}. It defines the
scattering in the model and leads eventually to
 the relation similar to (\ref{sc}) for the ZSF:
\beql{w6}
W_{\nu}(r;\mu)=\sum_{s\in W}b(s;\nu)V(r;s\nu,\mu).
\eq
In contrast to (\ref{sc}) it is the exact formula, although
the right hand side is
the sum of unbounded functions.
The calculation of the coefficients $b(s;\nu)$ is based on the following
important property of the WF (compare with (\ref{g15a}) for the ZSF)
\beql{w7}
W_{\nu}(r;\mu)=M(s;\nu,\mu)W_{s\nu}(r;\mu),
\eq
where $M(s;\nu,\mu)$ is a meromorphic function of $\nu$ which is determined
recursively as follows. If $s_{\al},~\al\in \Pi$ is a simple reflection, then
\beql{w8}
M(s_{\al};\nu,\mu)=e_{\al}(\nu)e_{\al}(-\nu)^{-1}
\left(\frac{|\mu_{\al}|}{\sqrt{2<\al,\al>}}\right)^{2\nu_{\al}},
\eq
$$
e_{\al}^{-1}(\nu)=\G\left(\frac{1}{2}-\nu_{\al}\right).
$$
Let $l(s)$ be the length of $s$. It means the minimal number of simple
reflections in $s=s_{\al_1}\ldots s_{\al_n}$. If $l(s_{\al}s)=l(s)+1$, then
\beql{w9}
M(s_{\al}s;\nu,\mu)=M(s;\nu,\mu)M(s_{\al};\nu,\mu).
\eq
These relations are coming from expressions for
the intertwiners for the equivalent representations $\pi_{\nu}$ and
$\pi_{s\nu}$. They
were investigated from this point of view in \cite{Sch}.

It can be proved that for $\phi\in\La^+,~\nu\in D$
$$\li V_{\nu}(e^{\phi t};\mu)e^{(s\nu+\rho)(\phi t)}=0,$$
unless $s=id$. Otherwise, this limit is equal to $1$.
In fact, $\Re e(\nu-s\nu)(\phi)<0$ for $s\neq id,~
\phi\in\La^+,~-\nu\in D$ (Lemma 3.3.2.1 in \cite{Wa}). Then, this statement
follows from the expansions of $\vf_{\nu}(\phi;\mu)$ (\ref{w2}).
We obtain from (\ref{w5}) $b(id;\nu)=c(-\nu)$. On the other hand,
it follows from (\ref{w6}) and (\ref{w7}) that
$$b( s_1s_2;\nu,\mu)=M(s_2;\nu,\mu)b(s_1;s_2\nu,\mu),$$
and, in particular,
\beql{w10}
b(s;\nu,\mu)=M(s;\nu,\mu)c(-s\nu).
\eq
Thus, eventually, the analog  of (\ref{sc}) for the WF takes the form
\beql{w11}
W_{\nu}(r;\mu)=\sum_{s\in W}M(s;\nu,\mu)c(-s\nu)V(r;s\nu,\mu),
\eq
where $M(s;\nu,\mu)$  is determined by (\ref{w8}) and
(\ref{w9}). We present expansion (\ref{w11}) explicitly determining
an appropriate normalization. For the rank one case
$V(r;s\nu,\mu)=\G(1/2-i\nu)I_{-i\nu}(\mu e^{-2\phi})$ (see (I.\ref{as}) and
(I.\ref{LWFI})).
We can multiply the WF by arbitrary function depending on $\nu$ only.
It still satisfies i), ii). We demonstrate that after
the transformation
 $$\ti{W}_{\nu}(r;\mu)=\xi(\nu)W_{\nu}(r;\mu),$$
where
\beql{w13}
\xi(\nu)=\prod_{\al\in\De^+}\frac
{\sin i\pi\nu_{\al}}{\pi}
\G(1/2-i\nu_{\al}),
\eq
all the poles in the expansion of $W'_{\nu}(r;\mu)$ disappear:
$$
\ti{W}_{\nu}(r;\mu)=\sum_{s\in W}b'(s;\nu,\mu)V(r;s\nu,\mu),~
$$
\beql{w12}
b'(s;\nu,\mu)=\prod_{\al\in\De^+}\frac{\det s}{\G(1+(-si\nu)_{\al})}
\left(\frac{|\mu_{\al}|}{\sqrt{2<\al,\al>}}\right)^{2\sum_{\ga}i\nu_{\ga}}.
\eq
Here $\sum_{\ga}$ is taken over those of the simple roots $\ga_i$ which
contribute to the representation $s=s_{\ga_1}\ldots s_{\ga_n}$, and $\det s=\pm
1, \det s_{\al}=-1$. The proof of (\ref{w12}) is given in Appendix B.
Moreover, if we choose $|\mu_{\al}|=\sqrt{2<\al,\al>}$, then, as it follows
from (\ref{w13}), (\ref{w12}), $\ti{W}_{\nu}(r;\mu)$ becomes $W$ antisymmetric
in the $\nu$ variables. Thus, this normalization is fixed up to the
multiplication by $W$ symmetric polynomial in $\nu$.

Now we define another type of the bounded WF, which we call the Gauss Whittaker
functions. It satisfies i) and, instead of ii),

\bigskip

\noindent
ii$^G)~W^G_{\nu}(vgn;\mu^L,\mu^R)=
\psi_{\mu^L}(n)\psi_{\mu^R}(v)W^G_{\nu}(g;\mu^L,\mu^R),~v\in \bar{N},
n\in N$.

\bigskip

\noindent
It follows from the Gauss decomposition that it also lives in the Cartan
algebra ${\cal A}$.
In particular, it satisfies the second order differential equation (see
(\ref{g13}))
\beql{w14}
\left[e^{-2\rho(\phi)}\sum_{j=1}^{r}\p_{\phi_j}e^{2\rho(\phi)}\p_{\phi_j}-
\sum_{\al\in\Pi}\mu_L\mu_Re^{-\al(\phi)}\right]W^G_{\nu}(\phi;\mu^L,\mu^R)=\\=
(-\nu^2+\rho^2)W^G_{\nu}(\phi;\mu^L,\mu^R).
\eq
This equation coincides with (\ref{w0}) for the Iwasawa WF after the
redefinition
\beql{r14a}
\nu_I=2\nu_G,~~\mu_L\mu_R=2\mu_I^2,
\eq
where subscribes  $G$ and $L$ refer to the Gauss and Iwasawa cases.

We suggest that this two functions after the identification of their
parameters coincide up to the multiplication
by $W$ symmetric polynomial in $\nu$.
 To this end,
we define $W^G_{\nu}(g;\mu^L,\mu^R)$ explicitly.  We need the state
$<\Psi_L^{\mu_L}|$ covariant with respect to the $\pi_{\nu}(\bar{N})$:
$$<\Psi_L^{\mu_L}|\pi_{\nu}(v)=\psi_{\mu_L}(v)<\Psi_L^{\mu_L}|,~v\in\bar{N}.$$
This state can be read off from the explicit realization of
$\pi_{\nu}(\bar{N})$ (\ref{g10}).
Let $S\equiv s_0$ be the longest element of the Weyl group. It transforms the
set of positive roots into the set of negative roots. We keep the same notation
for the representative of $S\in W\sim M'/M$ in $M'\subset K$. Thus,
$\Ad_{S} N=\bar{N}$. Consider the state $<~\Psi^{\mu}_L~|=\exp i\mu_L(y)$,
covariant with respect to the $N$ action.  Then, the state
$<\Psi^{\mu_L}_L|\pi_{\nu}(S)$ is $\bar{N}$-covariant
 $$
<\Psi^{\mu_L}_L|\pi_{\nu}(S)\pi_{\nu}(v)=\psi_{\mu_L}(v)
<\Psi^{\mu_L}_L|\pi_{\nu}(S).
$$
Explicitly,
$$
<\Psi^{\mu_L}_L|\pi_{\nu}(S)=h_G^{i\nu-\rho}(yS)\exp
i\mu_L(\bar{n}'(yS)).
$$
 It allows one to write down the GLWF:
\beql{w15}
W^G_{\nu}(r;\mu^L,\mu^R)=r^{-i\nu-\rho}\int_Ndyh_G^{i\nu-\rho}(yS)
\exp i\{\mu_L(n'(yS))+\mu_R(r^{-1}yr)'\}.
\eq
This integral converges absolutely for $\nu\in{\cal A}^*,~\log r\in {\cal A}$,
although we cannot prove it in general case.
The problem is that, in contrast to the
$K$-invariant states, the Whittaker vectors do not belong to the $L^2$-space.
But the case of the $SL(N)$ group (see (I.\ref{res})) shows that the poles of
$h_G^{i\nu-\rho}(yS)$ are canceled by the zeroes of $\exp i\mu_L(n'(yS))$.
Thus, the GLWF is also a bounded holomorphic in $\phi$ and $\nu$
solution to the same equation. There is only one solution of such a type.
Therefore, up to the normalization, the functions coincide.

Taking the limit $r=r(t)=\exp tH,~ H\in\La^+,~t\rar +\infty$, we find for
$-\nu\in D$ (\ref{g16}) the new representation of the standard Harish-Chandra
function
\beql{w16}
c(-\nu)\sim \int_Ndyh_G^{i\nu-\rho}(yS)
\exp i\mu_L(n'(yS)).
\eq
Non-trivial fact is that this integral can be factorized. This can be
proved much along the line of \cite{GK}.

\section{Classical reductions}
\setcounter{equation}{0}

Here we reproduce the classical Hamiltonian description of the open
Toda model using the symplectic reductions based on the Iwasawa and the
Gauss representations. It allows
to write two kinds of actions on the "upstairs"
space. It is a classical counterpart of the two Whittaker models described
above. In principle, using the functional integral technique, one can
calculate the Whittaker wave functions. We will proceed in this way
for affine groups.

\subsection{Iwasawa reduction}

Consider the cotangent bundle $T^*G$ for the real split group $G$. It is
defined
by the pair $(Y,g),~Y\in{\cal G}^*,~g\in G$. There is the canonical
bi-invariant
symplectic form on $T^*G$
\beql{r1}
\om=\de Y(\de gg^{-1})
\eq
and the set of invariant commuting Hamiltonians
\beql{r2}
\frac{<Y^{d_k}>}{d_k}, k=1\ldots,r,
\eq
where $d_k=2,\ldots$ are invariants of ${\cal G}$ and  $Y^{d_k}$
are  polynomials on $U({\cal G}^*)$.
It is the upstairs Hamiltonian system.

Now consider the gauge transform by the group
$K\oplus N$ coming from the Iwasawa decomposition
\beql{r3}
g\ms kg, ~Y\ms kYk^{-1},~~k\in K
\eq
$$
g\ms gn,~Y\ms Y,~~n\in N.
$$
It defines two moment maps
\beql{r4}
\mu_k=Pr_{{\cal K}^*}~ Y,~~~\mu_n=Pr_{{\cal N}^*}~ g^{-1}Yg,
\eq
where $Pr$ means the orthogonal projection with respect to the Killing form.
In the Iwasawa representation, $g$ can be transform by (\ref{r3}) to the
Cartan subgroup $A$. Let $g=\exp\phi,~\phi\in{\cal A}=h_I$.
Assume that
\beql{r5}
\mu_k=Pr_{{\cal K}^*} ~Y=0,~~~\mu_n=Pr_{{\cal N}^*}~ g^{-1}Yg=\mu_I,
\eq
where $\mu_I=\sum_{\al\in\Pi}\mu_{\al}{\cal G}_{\al}$ is the same as in
(\ref{woa}).
The first relation simply means that $Y\in {\cal P}$.
The second one can be easily solved and we
remain with an undetermined Cartan
component. Denote it $p\in {\cal A}^*$. We come eventually to the expression
\beql{r6}
Y=p+\sum_{\al\in\Pi}\mu_{\al}e^{\al(\phi)}\left({\cal G}_{\al}+
{\cal G}_{-\al}\right).
\eq
The reduced symplectic form acquires the canonical form
$$\om^{red}=\de p\de\phi=\sum_{k=1}^r\de p_k\de\phi_k.$$
The reduced phase space $K\setminus \setminus T^*G//N$
has the dimension
$$2\dim G-2\dim K -2\dim N=2r.$$
The combination $W(Y;\tau_1\ldots,\tau_r)=\sum_{k=1}^r\frac{\tau_k}{d_k}
<Y^{d_k}>$ defines the
classical hierarchy of the open Toda lattice.
In particular,
$$
<Y^2>=\frac{1}{2}\sum_{k=1}^rp_k^2+
\sum_{\al\in\Pi}\mu_{\al}e^{2\al(\phi)}
$$
 is the conventional Toda Hamiltonian, which after
the canonical quantization coincides with (\ref{w0}). The classical action
in this
representation takes the form
\beql{r7}
S^I=\int Y(\p_tgg^{-1})-<Y^2>+<B_K,Y>+<B_N,g^{-1}Yg-\mu_I>,
\eq
where $B_K\in {\cal K},~B_N\in {\cal N}$ are the Lagrange multipliers.

\subsection{Gauss reduction}

Consider the
symplectic reduction on $T^*G$ with $\om$ (\ref{r1}) under the action
of $\bar{N}\oplus N$
\beql{r8}
g\ms vg, ~Y\ms vYv^{-1},~~v\in \bar{N},
\eq
$$
g\ms gn,~Y\ms Y,~~n\in N.
$$
As previously, we have two moment maps
\beql{r9}
\mu_v=Pr_{\bar{\cal N}^*} ~Y,~~~\mu_n=Pr_{{\cal N}^*}~ g^{-1}Yg.
\eq
Using the Gauss decomposition, $g$ can be transform by (\ref{r8}) to the
Cartan subgroup $A$. Let $g=\exp\phi,~\phi\in{\cal A}$.
Assume that
\beql{r10}
\mu_v=Pr_{\bar{\cal N}^*} ~Y=\mu^L,~~~\mu_n=Pr_{{\cal N}^*} ~g^{-1}Yg=\mu^R,
\eq
where $\mu^{L}=\sum_{\al\in\Pi}\mu^{L}_{\al}{\cal G}_{\al}$,
$\mu^{R}=\sum_{\al\in\Pi}\mu^{R}_{\al}{\cal G}_{-\al}$ is the same as in
ii$^G$).
After "diagonalizing" $g$ by the Gauss representation at the point
$g=\exp\phi,~\phi\in{\cal A}=h_G$, we solve constraints (\ref{r10})
\beql{r11}
Y=p+\sum_{\al\in\Pi}\left(\mu^R_{\al}e^{\al(\phi)}{\cal G}_{\al}
+\mu^L_{\al}{\cal G}_{-\al}\right).
\eq
This representation of $Y$ differs from (\ref{r6}) by a Cartan gauge transform,
and, therefore, yields the  Hamiltonian which differs from (\ref{r14a})
by degrees in the potential. It corresponds to (\ref{w14}).
These two expressions of $Y$ are
just two different forms of the Lax
representation for the Toda lattice.

The classical action in this
representation takes the form
\beql{r12}
S^I=\int Y(\p_tgg^{-1})-<Y^2>+<B_L,Y-\mu^L>+<B_R,g^{-1}Yg-\mu^R>,
\eq
where $B_L\in \bar{\cal N},~B_R\in {\cal N}$ are the Lagrange multipliers.

\section*{Acknowledgments}
The authors are grateful to A.Zabrodin for useful discussions.
The work of A.G. is partially supported by RFFI-02-14365 and
INTAS-1010-CT93-0023, the work of S.K. and A.Mir. -- by RFFI-02-14365,
ISF-MGK300, INTAS-1010-CT93-0023 and by "Volkswagen Stiftung",
the work of A.Mar.  -- by RFFI-01-01106 and ISF-MGK300 and the work of M.O.
-- by ISF-MIF300, RFFI-01-01101 and INTAS-93-633. M.O. also thanks the
Max-Planck-Institut f\"{u}r Mathematik in Bonn for hospitality, where the
last part of the work was completed.

\section*{Appendix A}
\setcounter{equation}{0}
\def\theequation{A.\arabic{equation}}

$\Gamma$-function formulas used are \cite{GR,BMP,BE}
\be\label{AGamma}
\Gamma(z)={b^z\over\cos{\pi z\over 2}}\int_0^\infty t^{z-1}\cos bt\ dt,\\
B(x,y)=\int_0^\infty{t^{x-1}\over (1+t)^{x+y}}\ dt=
2\int_0^\infty {t^{2x-1}\over (1+t^2)^{x+y}}\ dt,\\
B(x,y)=\int_0^1(t^{x-1}(1-t)^{y-1}\ dt,\\
\cos\pi\nu={\pi\over\Gamma({\f 2}+\nu)\Gamma({\f 2}-\nu)},\\
\Gamma ({\f 2})=\sqrt{\pi},\\
\Gamma(2x)={2^{2x-1}\over\sqrt{\pi}}\Gamma (x)\Gamma (x+{\f 2}).
\ee
Cylindric function formulas used are
\be\label{AMac}
K_\nu(z)={\Gamma(\nu+{\f 2})\over\Gamma ({\f 2})}(2z)^\nu\int_0^\infty
{\cos t\over (t^2+z^2)^{\nu+{\f 2}}}\ dt,\\
K_\nu(z)={\f 2}\left({z\over 2}\right)^\nu\int_0^\infty
{e^{-t-{z^2\over 4t}}\over t^{\nu+1}}\ dt,\\
I_\nu(z)\stackreb{z\to 0}{\sim}{\f \Gamma (\nu+1)}\left({z\over
2}\right)^\nu.
\ee
Hypergeometric formulas used are
\be\label{hgir}
\phantom{F}_2F_1(a,b,c;z)={\Gamma(c)\over\Gamma(b)\Gamma(c-b)}
\int dtt^{b-1}(1-t)^{c-b-1}(1-tx)^{-a},
\ee
\be\label{hgpr}
\phantom{F}_2F_1(a,b;c;z)=(1-z)^{c-a-b}\phantom{F}_2F_1(c-a,c-b;c,z).
\ee
Integral representation for the Meyer function is
\be\label{irmf}
G^{mn}_{pq}\left(z\left|
\begin{array}{ccc}
a_1, & \ldots, & a_p\\
b_1, & \ldots, & b_q
\end{array}\right.\right)=\\=
{\f 2\pi i}\int {\Gamma(b_1+s)\ldots\Gamma(b_m+s)
\Gamma(1-a_1-s)\ldots\Gamma(1-a_n-s)\over\Gamma(a_{n+1}+s)\ldots\Gamma(a_p+s)
\Gamma(1-b_{m+1}-s)\ldots\Gamma(1-b_q-s)}\ z^{-s}ds.
\ee
This function can be also expressed through the finite sums of hypergeometric
functions \cite{BMP}.

\section*{Appendix B}
\setcounter{equation}{0}
\def\theequation{B.\arabic{equation}}
{\bf Proof of (\ref{w12})}\\

Let us prove first that $b(s;\nu,\mu)$ (II.\ref{w10}) can be represented as
\beql{a1}
b(s;\nu,\mu)=\frac{1}{\prod_{\al\in\De^+}\G(1/2-i\nu_{\al})}
\left(\frac{|\mu_{\al}|}{\sqrt{2<\al,\al>}}\right)^{2\sum_{\ga}-i\nu_{\ga}}
\prod_{\al\in\De^+}\G(-is\nu_{\al}).
\eq
Since $M(id;\nu,\mu)=1$, for $s=id$ (\ref{a1}) follows from
(II.\ref{g18}) and (II.\ref{w11}).
Assume that (\ref{a1}) is valid for all $s,~ l(s)\leq n$ and let us prove it
for $s_{\bet}s$. It follows from (II.\ref{w9}) that
\beql{a2}
b(s_{\bet}s;\nu,\mu)=\frac{\G(1/2-is_{\bet}(s\nu)_{\bet})}
{\G(1/2-i(s\nu)_{\bet})}
\left(\frac{|\mu_{\bet}|}{\sqrt{2<\bet,\bet>}}\right)^{-2i\nu_{\bet}}
M(s;\nu,\mu)c(-s_{\bet}s\nu).
\eq
Now
\beql{a3}
c(s_{\bet}s\nu)/c(s\nu)=\frac
{\G(1/2+(si\nu)_{\bet})\G(s_{\bet}(si\nu)_{\bet})}
{\G(1/2+s_{\bet}(si\nu)_{\bet})\G((si\nu)_{\bet})}.
\eq
This relation can be derived from the following statement:

\bigskip

\noindent
Each simple reflection $s_{\al}$ maps the subset of positive roots
$\De^+\setminus\al$ into itself
$$\bet-\frac{2<\al,\bet>}{<\al,\al>}\al\in\De^+,~
{\rm if~}\al\in\Pi,\bet\in\De^+,{\rm ~and~}\bet\neq\al.
$$

\bigskip

\noindent
Thereby, all the
survived gamma functions in the ratio $c(s_{\bet}s\nu)/c(s\nu)$
are represented in the right hand side
of (\ref{a3}). Substituting $c(s_{\bet}s\nu)$ in (\ref{a2}), we come to
(\ref{a1}).

It is clear that the function $f(\nu)=\prod_{\al\in\De^+}\sin\pi\nu_{\al}$ is
antisymmetric with respect to
the action of the Weyl group $f(s\nu)=\det  sf(\nu),~s\in W$.
Now from the relation
$$\frac{\sin \pi\nu_{\al}}{\pi}=[\G(1-\nu_{\al})\G(\nu_{\al})]^{-1}$$
we find that
$$\det  s\prod_{\al\in\De^+}\frac{i\sinh \pi\nu_{\al}}{\pi}=
\prod_{\al\in\De^+}[\G(1-(si\nu)_{\al})\G((si\nu)_{\al})]^{-1},$$
and, therefore,
$$\prod_{\al\in\De^+}\G(si\nu_{\al})=\frac{\det s}
{\prod_{\al\in\De^+}\G(1-(si\nu)_{\al})}
\prod_{\al\in\De^+}\frac{\pi}
{i\sinh \pi\nu_{\al}}.$$
Substituting this expression in (\ref{a1}), we find that
$$b(s;\nu,\mu)=b'(s;\nu,\mu)/\xi(\nu),$$
where $\xi(\nu)$ is common multiplier (II.\ref{w13}). Thereby,
(II.\ref{w12}) is proved.

\end{document}